\documentclass[12pt]{article}
\pdfoutput=1
\usepackage{ifpdf}
\usepackage[a4paper,top=3.0cm,width=16cm, left=2.6cm]{geometry}
\usepackage{graphicx}
\usepackage{microtype}
\usepackage[utf8x]{inputenc}
\usepackage[T1]{fontenc}
\usepackage{ae}
\usepackage{aecompl}
\usepackage{amssymb}
\usepackage{mathdots}

\usepackage[small,sf,bf,hang]{caption}
\usepackage[english]{babel}
\usepackage{datetime}
\shortdate 

\usepackage{authblk}

\title{{\bf Resurgence in Extended Hydrodynamics}}
\author[1]{In\^{e}s Aniceto}
\author[2,3]{Micha\l\ Spali\'nski}
\affil[1]{Institute of Physics, Jagiellonian University, ul. {\L}ojasiewicza 11, 30-348 Krak\'ow, Poland}
\affil[2]{National Center for Nuclear Research, ul. Ho\.za 69, 00-681 Warsaw, Poland}
\affil[3]{Physics Department, University of Bia{\l}ystok, Konstantego
  Cio\l kowskiego 1L, 15-245 Bia\l ystok, Poland}
\date{}

\long\def\symbolfootnote[#1]#2{\begingroup%
\def\thefootnote{\fnsymbol{footnote}}\footnote[#1]{#2}\endgroup}

\usepackage[nosort]{cite} 

\usepackage{collect}

\newcommand{\bea}{\begin{eqnarray}}
\newcommand{\beal}[1]{\begin{eqnarray}\label{#1}}
\newcommand{\eea}{\end{eqnarray}} 
\newcommand{\be}{\begin{equation}} 
\newcommand{\bel}[1]{\begin{equation}\label{#1}}
\newcommand{\ee}{\end{equation}} 
\newcommand{\rf}[1]{Eq.~(\ref{#1})}
\newcommand{\f}[2]{\frac{#1}{#2}}
\newcommand{\nn}{\nonumber}
\newcommand{\D}{{\mathcal D}}

\newcommand{\sym}{${\mathcal N}=4$}
\newcommand{\symm}{${\mathcal N}=4$ SYM}

\makeatother

\begin{document}

\maketitle

\thispagestyle{empty}

\begin{abstract}It has recently been understood that the hydrodynamic series
  generated by the M\"uller-Israel-Stewart theory is divergent, and that this
  large-order behaviour is consistent with the theory of
  resurgence. Furthermore, it was observed that the physical origin of this
  is the presence of a purely damped nonhydrodynamic mode. It is very
  interesting to ask whether this picture persists in cases where the spectrum
  of nonhydrodynamic modes is richer.  We take the first step in this
  direction by considering the simplest hydrodynamic theory which, instead of
  the purely damped mode, contains a pair of nonhydrodynamic modes of complex
  conjugate frequencies. This mimics the pattern of black brane quasinormal
  modes which appear on the gravity side of the AdS/CFT description of
  \symm\ plasma.  We find that the resulting hydrodynamic series is divergent
  in a way consistent with resurgence and precisely encodes information about
  the nonhydrodynamic modes of the theory.
\end{abstract}

\newpage

\section{Introduction}
\label{sec:intro}

Recent years have seen significant advances in the formulation of relativistic
hydrodynamic theories \cite{Baier:2007ix,Romatschke:2009im}. This is the
result of great interest in the heavy ion collision program, which aims to
establish bulk properties of nuclear matter in extreme conditions
\cite{Shuryak:2014zxa}. Relativistic hydrodynamic models have been essential
in uncovering the basic features observed in experiments at RHIC and LHC. As a
result, it has become clear that the hydrodynamic approach can be viewed in
the same spirit as the effective field theory paradigm of quantum field
theory. This has led to posing (and sometimes answering) foundational
questions about the meaning of hydrodynamics.

The point of departure is the idea that the expectation value of the
energy-momentum tensor can be expanded in gradients of hydrodynamic
variables. It has recently been shown in some specific cases that this
hydrodynamic gradient series is divergent
\cite{Heller:2013fn,Heller:2015dha}. Furthermore, the precise way in which the
series diverges encodes information about the nonhydrodynamic modes which are
not included explicitly in the hydrodynamic description. This pattern is
reminiscent of what has been noted in the context of divergent perturbation
expansions in other contexts.

The first example of such behaviour of the hydrodynamic gradient expansion was
the case of \sym\ supersymmetric Yang-Mills 
theory (SYM), where the series was calculated to high order using the
AdS/CFT correspondence \cite{Heller:2013fn}. To 
make the problem manageable, the specific case of boost-invariant 
flow \cite{Bjorken:1982qr,Janik:2005zt} was considered. The gradient expansion was computed up 
to order $242$, and it was observed that the series diverges in factorial
fashion. Applying the Borel transform revealed singularities in the Borel
plane occurring precisely at the locations corresponding to the complex
frequencies of the leading quasinormal modes of the dual black brane
geometry. The singularities are related by complex 
conjugation and are off the real axis, which means that the corresponding
degrees of freedom have oscillatory as well as decaying features. 

The second example where this kind of behaviour was observed is the
hydrodynamic series generated by M\"uller-Israel-Stewart (MIS) theory
\cite{Muller:1967zza,Israel:1979wp}. In this case the series is again
divergent, but the singularities in the Borel plane lie on the real axis,
which would correspond to purely decaying quasinormal modes. Since in this
example one has full control of the problem it is possible to resum the
series using ideas from the theory of resurgence \cite{Heller:2015dha}. This
example is very interesting from the point of view of Borel summation, since
the naive application of the inverse Borel transform leads to an imaginary
ambiguity. Proper accounting of the nonhydrodynamic degrees of freedom in a
way precisely consistent with resurgence theory yields a real and
unambiguous result 
(up to a constant of integration). 
This result was shown to
be consistent with an attractor 
solution in the original MIS equation, which constitutes a natural definition
of the meaning of hydrodynamics beyond the gradient expansion. This gives a
strong indication that in cases where a numerical solution is not available,
resurgence techniques may be used to mine the hydrodynamic gradient expansion
for universal features at times well before hydrodynamic behaviour is
typically expected.

The pattern seen in this problem is actually rather typical of the way
resurgence theory clarified the role and meaning of perturbative expansions in
physics.  The role of resurgence in the cancellation of ambiguities in quantum
mechanics is well known. In the perturbative study of observables such as the
ground state energy of the anharmonic oscillator
\cite{Bender:1969si,Bender:1990pd}, the observable studied was seen to be
"non-Borel summable" along the positive real line, since this was a so-called
Stokes line with singularities in the corresponding Borel plane. Nevertheless,
once all nonperturbative sectors associated with higher multi-instanton
corrections were taken into consideration, a process called median resummation
was seen to provide a real and unambiguous result (see, e.g.,
\cite{Bogomolny:1980ur,ZinnJustin:1981dx,ZinnJustin:1982td,ZinnJustin:1983nr,Jentschura:2004ib,Jentschura:2004cg,Ambrozinski:2012zw}
for examples of ambiguity cancelation in the context of quantum
mechanics,\footnote{It was seen in that in many quantum mechanical systems a
  very simple exact quantisation condition can be derived for the energy
  eigenvalues \cite{Dunne:2013ada,Dunne:2014bca}, relating perturbative and
  nonperturbative phenomena, which complements the usual large-order relations
  coming from resurgence.} and Refs. \cite{Argyres:2012ka,Dunne:2012ae} for its
generalisation to quantum field theory). This process consists in the proper
summation of all existing sectors (perturbative and nonperturbative) for a
given observable, in what is called a transseries. The use of median
resummation for the case of transseries with one and two real instanton
actions was studied in detail in Ref. \cite{Aniceto:2013fka}, and examples of
recent applications are Refs. \cite{Heller:2015dha} and
\cite{Aniceto:2015rua,Dorigoni:2015dha}.

The cancellation of nonperturbative ambiguities is just one application of a
much larger structure behind the asymptotic behaviour of perturbative
series. In fact, resurgent analysis and transseries give us a straightforward,
systematic path of determining the analytic properties of the observables, the
Stokes phenomena associated with singular directions, and the resummation
properties leading to unambiguous results.\footnote{See for example
  \cite{Candelpergher:1993np,Delabaere:2006ed,Seara:2003ss,sauzin14,Dorigoni:2014hea,Edgar:2008ga}
  for reviews on resurgent analysis and transseries, and
  \cite{Marino:2008ya,Aniceto:2011nu,Marino:2012zq,Upcoming:2015} for
  introductions to resurgence in physical settings. For recent applications to
  topological strings, supersymmetric quantum mechanics and QFTs, see also
  \cite{Santamaria:2013rua,Couso-Santamaria:2014iia,Hatsuda:2015owa,Couso-Santamaria:2015hva,Basar:2013eka,Behtash:2015kva,Cherman:2014ofa,Dunne:2015ywa}. }
The crucial role of resurgence in the study of the analytic properties and
Stokes phenomena within physical contexts is exemplified in
Ref. \cite{Couso-Santamaria:2015wga}, where from a large-$N$ expansion one can 
retrieve the properties of the corresponding transseries solution not only for
real, finite $N$, but also as an analytic function in the variable $N$ (see
also Ref. \cite{Cherman:2014xia} for another toy example of strong-weak coupling
interpolation).

One can ask moreover about the usefulness of the transseries and resurgence in
cases where the resummation procedure in the direction of interest is not
singular.
 When our interest is in the result along the positive real line,
and the singular directions (Stokes lines) are in the complex plane away from
this axis, one could be led to believe that only the perturbative
series would be necessary. But as it is known from resurgent theory, and
evidence was seen, for example, in Ref. \cite{Grassi:2014cla}, the existence of
Stokes lines with a positive real component will introduce nonperturbative
sectors which need to be added to the original asymptotic perturbative series
in the form of a transseries in order to obtain a consistent
  result. 

It would clearly be interesting to apply the ideas of resurgence theory to the
case of \sym\ SYM. Applications of resurgence for supersymmetric gauge
theories were already seen in the case of localisable observables
\cite{Aniceto:2014hoa} and relations to quantum mechanical systems
\cite{Basar:2015xna}.  Here we turn to a hydrodynamic model which shares some
of the simplicity of MIS theory, but contains a richer spectrum of
nonhydrodynamic modes in a way which resembles some aspects of what is known
about \sym\ SYM.

It is important to recall here the
  philosophy behind Ref. \cite{Heller:2015dha}: models like MIS are regarded as
  means of generating the hydrodynamic gradient expansion which is then
  analysed as if it came from a microscopic theory. 
Specifically, we will
study a hydrodynamic theory which 
contains analogs of quasinormal modes whose frequencies possess both real and
imaginary parts, as is the case for \sym\ SYM (and unlike MIS). In such a case
one would expect that the singularities of the Borel transform would be off the
real axis. The simplest
such example is one of the models put forth in \cite{Heller:2014wfa}, where
nonhydrodynamic modes corresponding to quasinormal modes of  \sym\ SYM were
incorporated into a MIS-like theory. 
This model generates the same hydrodynamic expansion
as MIS theory up to second order in gradients (higher orders differ, of
course).
We show that also in this 
case one can identify attractor behaviour which sets in well before the
hydrodynamic limit of large times.  

We study the hydrodynamic series in this model in the spirit of
Ref.~\cite{Heller:2015dha} and find a similar picture, albeit with novel 
elements. The hydrodynamic series is divergent and its summation requires 
exponentially suppressed corrections reflecting in a quantitative manner the
spectrum of nonhydrodynamic modes present. These exponential corrections to
the hydrodynamic series can be viewed as a completion to a transseries. By
using the formalism 
elaborated
 in \cite{Aniceto:2011nu} we show that the
divergent series satisfy relations expected on the basis of resurgence
theory. From this perspective there is a novel aspect: the ``actions'' are
complex, as is the leading nonanaliticity exponent. This introduces some
technical difficulties in applying convergence acceleration. The physical
reason for these features is, however, entirely clear: they correspond to the
fact that the nonhydrodynamic modes present in this theory are not purely
decaying (i.e., the quasinormal mode frequencies are not purely imaginary).

The structure of this paper is as follows. We start by reviewing the important
aspects of hydrodynamic theories in Section~\ref{sec:hydro}, followed by more specific
properties of the MIS causal hydrodynamic theory in
Section~\ref{sec:mis}. Section~\ref{sec:misres} 
then presents the natural contact between resurgence and the ambiguity
cancellation of the MIS theory (reviewing the results of
Ref.~\cite{Heller:2015dha} in light of
Refs.~\cite{Aniceto:2011nu,Aniceto:2013fka}) as a warmup example toward the
extended theories of hydrodynamics.  
Section~\ref{sec:ext} introduces the hydrodynamic model which we will
  consider in the main part of this article.
The application of resurgence techniques to this theory is the main focus of
our work and is described in Section~\ref{sec:extres}. We will close with a
brief summary and ideas for the future in Section~\ref{sec:sum}.

\section{Hydrodynamics}
\label{sec:hydro}

Phenomenological equations of hydrodynamics are designed to
  reproduce the gradient expansion of the energy-momentum tensor in some
  microscopic theory up to some order (typically 1 or 2). The evolution
  equations are the conservation equations 
\bel{cons}
\nabla_\mu T^{\mu\nu} = 0
\ee
of the energy-momentum tensor
  expressed in terms of the {\em hydrodynamics variables}. Specifically, the
  energy-momentum tensor in the hydrodynamic theories considered here can be
  presented as
\bel{hydro}
T^{\mu \nu} = {\cal E} \,  u^{\mu} u^{\nu} +  {\cal P} ({\cal E}) (
\eta^{\mu \nu} + u^{\mu} u^{\nu} ) + \Pi^{\mu \nu},
\ee
where $\Pi^{\mu \nu}$ is the shear stress tensor (discussed in detail below), 
${\cal E}$ is the energy density and $ {\cal P}$ is the pressure, expressed
in terms of the energy density by an assumed equation of state. In conformal
theories in $d=4$ dimensions it takes the form
\be
{\cal P} ({\cal E})=\f{1}{3} {\cal E} \, .
\ee
The energy density ${\cal E}$ is often expressed in terms of the
``effective temperature'' $T\sim{\cal E}^{1/4}$. The field $u$ is the flow
velocity, defined as a timelike eigenvector of the energy-momentum tensor.
The spacetime dependent energy density (or effective temperature) and flow
velocity are the hydrodynamic variables, the evolution of which one wishes to
describe.  
Their precise definition away from equilibrium is what constitutes a choice of
hydrodynamic frame (see, e.g., Ref.~\cite{Bhattacharya:2011tra}). We adopt the
Landau frame, which means that 
we impose the condition that the shear stress tensor is transverse to the
flow: 
\be
u_\mu \Pi^{\mu \nu} = 0\, .
\ee
The hydrodynamic gradient expansion
is the approximation of $\Pi^{\mu \nu}$ by a series of terms, graded
by the number of spacetime gradients of the hydrodynamic fields $u^\mu$ and $T$. 

To proceed, it is highly advantageous to exploit to the fullest the constraints 
imposed by conformal symmetry. This desire has led to the development of the
so-called Weyl-covariant  
formulation \cite{Loganayagam:2008is} of conformal relativistic hydrodynamics,
in which the evolution equations assume a very compact form. 
We will not review this formalism here, but we will mention some of
  its basic features. The essential idea is to introduce a (nondynamical)
  ``Weyl connection''
\be
{\cal A}_{\mu} = u^{\lambda} \nabla_{\lambda} u_{\mu} - \frac{1}{3}
\nabla_{\lambda} u^{\lambda} u_{\mu} \, ,
\ee
to define a derivative operator, denoted here by
$\D_\mu$, which is covariant under Weyl transformations (spacetime dependent
rescalings) of the metric. 
The action of the Weyl-covariant derivative depends on the tensor on which it
acts. A general formula can be found in Ref.~\cite{Loganayagam:2008is}. 

It will also be convenient to define 
\be
\D\equiv u^{\mu} \D_{\mu} \ .
\ee
and 
\be
 \sigma^{\mu\nu} = \D^\mu u^\nu + \D^\nu u^\mu , \qquad \omega^{\mu\nu} =
 \D^\mu u^\nu - \D^\nu u^\mu \, .
\ee
These objects are transverse and transform homogeneously under Weyl
transformations~\cite{Loganayagam:2008is}. 

The Landau-Lifschitz formulation of relativistic viscous
hydrodynamics~\cite{LLfluid} asserts that 
\bel{PiLL}
\Pi^{\mu \nu} = - \eta \sigma^{\mu \nu} \, ,
\ee
where $\eta$ is the shear viscosity. 
Unfortunately, the resulting theory does not have a well-posed initial value
problem due to superluminal signal 
propagation~\cite{Hiscock:1985zz,PhysRevD.62.023003}. 
The same problem will occur if on the right-hand side of \rf{PiLL} one
includes any finite number of terms graded 
by the number of derivatives of $T$ and $u^{\mu}$. In principle these
problems appear at short distances, where hydrodynamics is not expected to
apply~\cite{Geroch:1995bx,Geroch:2001xs}, but for practical applications this
is no consolation because 
acausality leads to numerical instabilities. For practical
purposes it is therefore necessary to replace \rf{PiLL} by a prescription
which effectively generates all orders in the gradient expansion.

\section{MIS causal hydrodynamics}
\label{sec:mis}

MIS theory resolves the causality problem by promoting the shear stress tensor
$\Pi^{\mu\nu}$ to an independent dynamical field which satisfies a relaxation
type differential equation \cite{Muller:1967zza,Israel:1979wp} chosen to
augment the conservation law \rf{cons}. Consistency with the gradient
expansion requires that terms of at least second order be included, since the
derivative of the shear stress tensor is of second order.

If all terms admitted by symmetry are incorporated, the leading
terms in the gradient expansion of the shear stress tensor can be written as
\bel{st2ord}
  \Pi^{\mu\nu} = -\eta \sigma^{\mu\nu} + \eta\tau_\Pi \D\sigma^{\mu\nu} +
  \lambda_1 {\sigma^{<\mu}}_\lambda \sigma^{\nu>\lambda} + \lambda_2
         {\sigma^{<\mu}}_\lambda \omega^{\nu>\lambda} + \lambda_3
         {\omega^{<\mu}}_\lambda \omega^{\nu>\lambda} \, ,
\ee
where $<\dots>$ denotes symmetrization and subtracting the trace, and
$\tau_{\Pi}$ and $\lambda_i$ are phenomenological
parameters~\cite{Baier:2007ix} (the second-order transport coefficients).
If the energy-momentum tensor is calculated in some microscopic conformal
theory and expressed in terms of the hydrodynamic variables up to second order
in gradients, the result will be of the form of \rf{st2ord} with
  some specific values of the transport coefficients. It has, for example,
been obtained as the long-wavelength effective description of strongly coupled
${\cal N} = 4$ SYM plasma in the framework of the AdS/CFT
correspondence~\cite{Janik:2005zt,Heller:2007qt,Baier:2007ix,Bhattacharyya:2008jc}.

The main idea of treating hydrodynamics as an effective theory is to write
down an evolution equation, the gradient expansion of which generates
\rf{st2ord}, together with additional terms which 
are of third order and above. This can be done by eliminating
$\sigma^{\nu\lambda}$ in the second-order terms in favor of $\Pi^{\mu\nu}$
using \rf{PiLL}. The result can be written as  
\bel{mis2}
(\tau_\Pi \D + 1) \Pi^{\mu\nu} =  -\eta \sigma^{\mu\nu} +
\f{\lambda_1}{\eta^2} {\Pi^{<\mu}}_\lambda \Pi^{\nu>\lambda}
- \f{\lambda_2}{\eta} 
       {\Pi^{<\mu}}_\lambda \omega^{\nu>\lambda} + \lambda_3
       {\omega^{<\mu}}_\lambda \omega^{\nu>\lambda} \, .
\ee
Solving this iteratively yields \rf{st2ord} up to higher-order terms, as
desired. The coefficients of these higher-order terms will 
all be expressed in terms of the second-order transport coefficients which appear
explicitly in \rf{mis2}.

Linearization of the resulting theory reveals a single, purely decaying,
nonhydrodynamic mode in 
addition to hydrodynamic modes \cite{Baier:2007ix}. 
This mode (which we refer to as the MIS mode) decays on a scale set by $\tau_\Pi$.
Furthermore, the resulting theory  is causal as long as $T \tau_{\Pi}\geq \eta/s$. 
This approach has enjoyed great success in describing the evolution of
quark-gluon plasma~\cite{Luzum:2008cw}.  

In Ref.~\cite{Heller:2015dha}, the special case of Bjorken
flow~\cite{Bjorken:1982qr} was 
considered, and we do the same in our work. 
Due to a very high degree of symmetry imposed, the hydrodynamic equations 
reduce to a set
of ordinary differential equations. 
The symmetry in question, boost invariance, can be taken to mean that in
proper time-rapidity coordinates  
$\tau, y$ (related to Minkowski coordinates $t, z$ by the relations $t = \tau
\cosh y$ and 
$z = \tau \sinh y$) the energy density, flow velocity and shear stress tensor
depend only on the proper time $\tau$. The MIS equations \rf{mis2} then
reduce to 
\beal{miseqn}
\tau  \dot{\epsilon} &=& - \frac{4}{3}\epsilon + \phi\nonumber\, , \\
\tau_\Pi \dot{\phi} &=& 
\frac{4 \eta}{3 \tau } 
- \frac{\lambda_1\phi^2}{2 \eta^2}
- \frac{4 \tau_\Pi\phi}{3 \tau }
- \phi \, ,
\eea
where the dot denotes a proper time derivative 
and $\phi\equiv-\Pi^{y}_{y}$,
the single independent component of the shear stress 
tensor. 

In a conformal theory $\epsilon \sim T^4$ and 
the transport coefficients satisfy
\bel{contra}
\tau_\Pi = \frac{ C_{\tau \Pi }}{T}, \qquad \lambda_1 =  C_{\lambda_1}
\frac{\eta}{T}, \qquad \eta = C_\eta\ s \, ,
\ee
where $s$ is the entropy density and $C_{\tau \Pi }, C_{\lambda_1}, C_\eta$ are
dimensionless constants. In the case of  
\symm\ their values are known from fluid-gravity duality~\cite{Bhattacharyya:2008jc}:
\bel{symvalues}
 C_{\tau \Pi } = \frac{2-\log (2)}{2 \pi} , \qquad  C_{\lambda_1} = \frac{1}{2
     \pi}, \qquad  C_\eta = \frac{1}{4 \pi} \, .
\ee
To simplify the discussion we will consider the case $C_{\lambda_1} = 0$. 
This choice does not modify the nonydrodynamic sector in a
qualitative way, so our study of resurgence is not
affected. The hydrodynamic theory still matches \symm\ at the
level of first-order (viscous) hydro, which is physically by far the most
significant point.

Using \rf{contra} one can turn the system of equations \rf{miseqn} into a
single second-order differential equation for the proper time
dependence of the temperature $T(\tau)$.  
It proves fruitful to introduce the dimensionless variables
\bel{wfdef}
w =T\tau, \quad f = \f{\tau}{w}\f{dw}{d\tau} \, .
\ee
In terms of these, the second-order ordinary differential equation for $T(\tau)$
implies a first-order equation for $f(w)$, 
\bel{eq:first-order-MIS}
C_{\tau\Pi}f\,f' +
4C_{\tau\Pi}f^{2}+\left(w-\frac{16C_{\tau\Pi}}{3}\right)f - \frac{4C_{\eta}}{9}
+ \frac{16C_{\tau\Pi}}{9}-\frac{2w}{3}=0 \, , 
\ee
where $f'$ stands for the derivative of $f(w)$ with respect to $w$. 
It is this equation which is the starting point for our analysis of MIS
theory.

The late proper time behaviour of the system is governed by hydrodynamics. In
terms of the dimensionless variable $w$ this translates to the limit
$w\rightarrow\infty$. One can easily determine the coefficients of the series
solution valid for large $w$:
\bel{mishydro}
f(w) = \f{2}{3} + \frac{4 C_\eta}{9 w}+\frac{8 C_\eta C_{\tau \Pi}}{27
   w^2}+ O(\frac{1}{w^3}) \, .
\ee
This expansion corresponds to the hydrodynamic gradient expansion
\cite{Heller:2011ju}. By examining the behaviour of the coefficients in
\rf{mishydro} one can show that the series as divergent.  This fact reflects
the presence of the nonhydrodynamic MIS mode. As shown in
\cite{Heller:2015dha} the series solution can be summed using Borel techniques
by incorporating exponential corrections to the hydrodynamic expansion. The
result is a transseries, as described in detail in the following section.

\section{Resurgence and ambiguity cancellation}
\label{sec:misres}

As an introduction to the methods of resurgence theory we first
consider the case of MIS reviewed in the previous section. Unlike the analysis
presented in Ref.~\cite{Heller:2015dha}, we will not discuss ambiguity cancellation
at the level of the analytic continuation of the Borel transform. 
We will instead make use of the consistency conditions derived from alien
calculus (for a recent review and derivations of the formulae used in this
Section see Refs.~\cite{Aniceto:2011nu,Dorigoni:2014hea,Upcoming:2015}). 
As was already seen in Ref.~\cite{Heller:2015dha}, this
example presents some remarkable resurgent properties, while being a
very simple application of the expressions derived in
Ref.~\cite{Aniceto:2011nu,Aniceto:2013fka}. 

In order to capture the full 
solution to the first-order differential equation \rf{eq:first-order-MIS} from
a perturbative expansion,  
one needs a transseries Ansatz with one parameter 
\be
f(w,\sigma) = \sum_{n=0}^{+\infty} \sigma^{n} \mathrm{e}^{-nAw}
\Phi_{n} \left(w\right) \, ,
\ee
where $\sigma$ is a parameter to be fixed by the physical properties
of our system -- in our case these are reality and initial
conditions. The constant $A$ is often referred to as the
instanton action, due to its interpretation in applications to perturbation
expansions in quantum field theory \cite{Bogomolny:1980ur,ZinnJustin:1981dx,ZinnJustin:1982td,ZinnJustin:1983nr,Jentschura:2004ib,Jentschura:2004cg,Ambrozinski:2012zw}. 

The $\Phi_{n}\left(w\right)$ are perturbative 
expansions around the nonperturbative, exponentially suppressed sectors
with contributions weighted by $\mathrm{e}^{-nAw}$
\begin{equation}
\Phi_{n}\left(w\right)=w^{\beta_{n}}\sum_{k=0}^{+\infty}a_{k}^{(n)}w^{-k}
\, . 
\label{eq:trans-series-one-param}
\end{equation}
The \textquotedbl{}instanton\textquotedbl{} action $A$ and coefficients
$\beta_{n}$ (the first associated 
with
the position of the cuts appearing
in the Borel plane, and the latter associated with the type of these
branch cuts) were determined in Ref.~\cite{Heller:2015dha} to be 
\be
A=\frac{3}{2}C_{\tau\Pi}, \qquad \beta_{n}\equiv
n\beta=-n\frac{C_{\eta}}{C_{\tau\Pi}} \, .
\ee
These coefficients, as well as the perturbative coefficients $a_{k}^{(n)}$
can be determined iteratively by substituting the transseries Ansatz
(\ref{eq:trans-series-one-param}) into the differential equation
(\ref{eq:first-order-MIS}). For the perturbative series $\Phi_0 (w)$ this
leads to   
\beal{eq:mishydro}
a_{0}^{(0)} &=& \f{2}{3} \nn\\
a_{1}^{(0)} &=& \f{4  C_\eta}{9}\nn\\
a_{k+1}^{(0)} &=& C_{\tau\Pi} \left(\f{16}{3} a_{k}^{(0)} - \sum_{n=0}^{k}
(4-n) a_{k-n}^{(0)} a_{n}^{(0)}  \right) ,\quad k>1 \, .
\eea
%
This recursion relation
makes it manifest that the series is indeed divergent.  

The expansions $\Phi_{n}\left(w\right)$ are asymptotic, and
the coefficients $a_{k}^{(n)}$ were seen to grow factorially for
large enough order $k$. The corresponding Borel transforms, schematically
of the form\footnote{The rule is to substitute $w^{-\alpha}\rightarrow s^{\alpha-1}/\Gamma\left(\alpha\right)$,
but one needs to remove any initial terms with $\alpha<0$, and add
them separately at a later stage: hence the $m_{\mathrm{min}}$ introduced in
\rf{eq:Borel-one-param-sectors}. This does not change the 
asymptotic nature of the series.} 
\be
\mathcal{B}\left[\Phi_{n}\right]\left(s\right)=\sum_{m=m_{\mathrm{min}}}^{+\infty}a_{m}^{(n)}
\frac{s^{m-n\beta-1}}{\Gamma\left(m-n\beta\right)} \, , 
\label{eq:Borel-one-param-sectors}
\ee
have a nonzero radius of convergence. The radius of convergence is in fact
given by the position of the first branch cut in the Borel plane, which is at
a distance $s=A$. Following the analysis presented in Ref.~\cite{Aniceto:2011nu}
(see Section 2 of this paper for more details), we know that in the case of a
one-parameter transseries with real positive 
instanton action $A$, the
sectors $\Phi_{0},\, \Phi_{1}$ will have cuts starting at positions $s=\ell A$
for $\ell \ge 1$ in the positive real axis, while the sectors $\Phi_{n}$ with
$n \ge 2$ will have cuts both in the negative and positive real lines on the
Borel plane: a finite number in the negative real axis at $s=\ell_{1} A$ with
$\ell_{1}=1,\cdots,n-1$, and an infinite number in the positive real axis at
$s=\ell_{2}A$ with $\ell_{2}\ge1$.

Given the Borel transforms, \rf{eq:Borel-one-param-sectors}, one can use
suitable Pad\'{e} approximants (see Ref.~\cite{Aniceto:2011nu}) and resum each
sector via the Laplace transform. The resummation can be easily performed in
directions $\theta$ in the complex plane where the Borel transforms
$\mathcal{B}\left[\Phi_{n}\right]$  do not have singular behaviour:
\begin{equation}
\mathcal{S}_{\theta}\Phi_{n}\left(w\right)=\int_{0}^{+\infty\,\mathrm{e}^{\mathrm{i}\theta}}ds\,
\mathrm{e}^{-sw} \mathcal{B}\left[\Phi_{n}\right] \left(s\right) \, .
\label{eq:resummation-one-param-sector}
\end{equation}
\noindent
This resummation can then be trivially analytically continued up to the
singular directions in the Borel plane, also known as Stokes lines. The
resummed transseries can then be defined by 
\begin{equation}
\mathcal{S}_{\theta}f\left(w,\sigma\right)=\sum_{n=0}^{+\infty}
\sigma^{n}\mathrm{e}^{-nAw} \mathcal{S}_{\theta} \Phi_{n}\left(w\right) \, . 
\label{eq:resummed-one-param-trans-series} 
\end{equation}

The transseries parameter $\sigma$ is free at this stage. Its (in general
complex) value can be determined by  
enforcing some physical constraints on the transseries --  in this case
suitable initial and reality conditions for \rf{vcp}. 

For physical reasons, we are ultimately interested in real and positive values
of the expansion parameter $w$. For the particular 
case in this Section, we have one further problem: the Borel transforms
of the perturbative and nonperturbative sectors have branch cuts
in the positive real axis starting at positions $s=\ell A$. This means
that the positive real axis is a Stokes line, a singular direction
in the complex plane where 
the Stokes phenomenon occurs. 
The Laplace transform
in (\ref{eq:resummation-one-param-sector}) is ill defined, because
of these branch cuts, and we have to define lateral resummations by
avoiding these singularities either from above or from below the real
axis:
\begin{equation}
\mathcal{S}_{\pm}\Phi_{n}\left(w\right)=\int_{0}^{+\infty\mathrm{e}^{\pm\mathrm{i}\epsilon}}
ds\,\mathrm{e}^{-sw}\mathcal{B}\left[\Phi_{n}\right]\left(s\right) \, .
\label{eq:lateral-resummations} 
\end{equation}
This introduces a nonperturbative ambiguity: if the coefficients $a_{m}^{n}$ are real, then the difference between
these two lateral resummations for each sector is pure imaginary and of
the order of $\mathrm{e}^{-Aw}$. For example, starting with
the lateral resummation above the real axis, we can define its real
and imaginary contribution
\be
\mathcal{S}_{+}\Phi=\frac{1}{2}\left(\mathcal{S}_{+}+\mathcal{S}_{-}\right)
\Phi +\frac{1}{2} \left(\mathcal{S}_{+}-\mathcal{S}_{-}\right) \Phi \equiv
\mathcal{S}_{R}\Phi + \mathrm{i}\mathcal{S}_{I}\Phi \, ,
\ee
where $\mathcal{S}_{I}\Phi_{n}\left(w\right)\sim\mathrm{e}^{-Aw}$. 
 One is interested in having a nonambiguous real transseries solution,
i.e., a transseries of the type (\ref{eq:resummed-one-param-trans-series})
but where now the resummation should be thought of as one of the lateral
resummations $\mathcal{S}\rightarrow\mathcal{S}_{\pm}$. Because $\sigma$
is a complex number, the real and imaginary contributions from each
sector $\mathcal{S}_{R,I}\Phi_{n}$ will mix with the real and imaginary
part of the parameter $\sigma\equiv\sigma_{R}+\mathrm{i}\sigma_{I}$,
and one can determine the total real and imaginary contributions to
the transseries coming from every sector. This was done in detail in \cite{Aniceto:2013fka}. 

Due to the resurgent properties of the transseries, choosing the parameter
$\sigma$ properly will cancel the imaginary \textquotedbl{}ambiguous\textquotedbl{}
contribution to the transseries, leaving us with a nonambiguous real
result. This procedure coincides with the so-called median resummation.
In \cite{Aniceto:2013fka} it was seen that for a one-parameter transseries with real
coefficients and singularities in the Borel plane lying in the positive
real axis,\footnote{Recall that higher nonperturbative sectors $\Phi_n$ with
  $n\ge2$ will naturally have singularities also in the negative real line, as
  it is expected from a one-parameter transseries. This is a natural part of
  the resurgent analysis and its consequences are already integrated in the
  analysis presented in Refs.~\cite{Aniceto:2011nu,Aniceto:2013fka}.} the median
resummation was achieved by setting the imaginary 
part of the transseries parameter to 
\be
\mathrm{i}\sigma_{I}=-\frac{1}{2}S_{1} \, .
\ee
Here $S_{1}$ is the so-called Stokes constant associated with the Stokes transition
across the positive real axis. The real part of the parameter $\sigma$
does not get fixed by these requirements, and 
remains as an integration constant, to be fixed by some initial condition. 
The nonambiguous real transseries
solution to this problem is therefore given by
\be
\mathcal{S}_{R}f=\mathcal{S}_{+}f\left(w,\sigma_{R}-\frac{1}{2}S_{1}\right)
= \sum_{n=0}^{+\infty}\left(\sigma_{R}-\frac{1}{2}
S_{1}\right)^{n}\mathrm{e}^{-nAw}\mathcal{S}_{+} \Phi_{n}\left(w\right) \, . 
\ee

The Stokes constant can be determined directly by using resurgence
formulae predictions for the large-order behaviour of the perturbative
series (as well as higher sectors).
From the resurgent analysis of the one-parameter transseries, it was seen
in Ref.~\cite{Aniceto:2011nu} that the discontinuity of the sectors $\Phi_{k}$
in the positive real direction $w=\left|w\right|\mathrm{e}^{\mathrm{i}\theta}$
with $\theta=0$ is given by
\begin{equation}
\mathrm{Disc}_{0}\Phi_{k}=-\sum_{\ell=1}^{+\infty}
\frac{(k+\ell)!}{k!\,\ell!}\left(S_{1}\right)^{\ell} \mathrm{e}^{-\ell
  Aw}\Phi_{k+\ell} \left(w\right) \, .
\label{eq:discontinuity-zero-one-param-sectors}
\end{equation}
For the particular case of the perturbative expansion we need only
to set $k=0$. For $k\ge2$ the sectors $\Phi_k$ will also have discontinuities
in the direction $\theta=\pi$. The full expressions for these discontinuities
were also derived in Ref.~\cite{Aniceto:2011nu} (in Section 2), but given the
length of such expressions we refer the reader to that reference for more
details. The fact that our transseries is resurgent translates 
directly to the existence of large-order relations between the coefficients
$a_{k}^{(n)}$ and $a_{k'}^{(m)}$ of neighbouring sectors $n,m$.
These large-order relations can be derived using Cauchy's theorem, 
\begin{equation}
F\left(w\right)=\oint_{z=w}\frac{dz}{2\pi\mathrm{i}}\frac{F\left(z\right)}{z-w}
= \sum_{\theta-\mathrm{sing}}\int_{0}^{+\infty\mathrm{e}^{\mathrm{i}\theta}}
\frac{dz}{2\pi\mathrm{i}} \frac{\mathrm{Disc}_{\theta}F\left(z\right)}{z-w} +
\oint_{\infty} \frac{dz}{2\pi\mathrm{i}}\frac{F\left(z\right)}{z-w} \, ,
\label{eq:Cauchy-thm}
\end{equation}
where the sum is over all singular directions $\theta$ of the asymptotic
expansion $F(w)$. On the rhs, we have deformed the contour of integration to
encircle  all the discontinuities (associated with the
singular directions $\theta$) and the contribution at infinity. 
Under certain conditions \cite{Bender:1990pd,Collins:1977dw}, 
the contribution at infinity can be  
seen to vanish by scaling arguments and we are left with the integration over
the discontinuities of $F\left(w\right)$. 

As an example choose the perturbative sector $\Phi_{0}$ and use variables
$x=w^{-1}\ll1$. We can easily write  
\be
x\Phi_{0}\left(x\right) = \oint_{z=x}\frac{dz}{2\pi\mathrm{i}} 
\frac{z\Phi_{0}\left(z\right)}{z-x} = \int_{0}^{+\infty}
\frac{dz}{2\pi\mathrm{i}} \frac{z\,\mathrm{Disc}_{0}\Phi_{0}(z)}{z-x} \, . 
\ee
In the above formula we used that the perturbative series has only
a discontinuity in the direction $\theta=0$.\footnote{Recall again that for
  $F(w)=\Phi_{k}(w)$ with $k=0,1$ we have only one singular direction
  $\theta=0$, while for $k\ge2$ we have two singular directions
  $\theta=0,\pi$.} We can now use
(\ref{eq:discontinuity-zero-one-param-sectors}) 
and expand both sides for small $x$. Comparing equal powers of $x$
we find
\be
a_{m}^{(0)}\simeq-\sum_{k=1}^{+\infty}
\frac{\left(S_{1}\right)^{k}}{2\pi\mathrm{i}}
\frac{\Gamma\left(m+k\beta\right)}{(kA)^{m+k\beta}}
\sum_{h=0}^{+\infty}a_{h}^{(k)} \prod_{\ell=1}^{h}
\frac{k\,A}{(m+k\beta-\ell)},\:m\gg1 \, .
\ee
This is a large-order relation with connects coefficients of the perturbative
series $a_{m}^{(0)}$ for large-order $m$ with coefficients of the
one-instanton series $a_{h}^{(1)}$ at low order, up to contributions of the
two-instanton series $a_{h}^{(2)}$ exponentially
suppressed by $2^{-m}$, and so on.  
Note that each of the sums appearing above is asymptotic. Similar
expressions have been derived for the coefficients of the other sectors, and
can be found in Ref.~\cite{Aniceto:2011nu}. For the coefficients of the higher
sectors $\Phi_k,\, k\ge2$ other Stokes constants will appear in the
large-order relations, because of the discontinuity in the direction
$\theta=\pi$ 
(see Section 2 of Ref.~\cite{Aniceto:2011nu} for more details).  

Returning to the
perturbative series, we can write the large-order relations  more explicitly
in the following way:  
\bea
\label{eq:ratio-one-param-pert}
\frac{2\pi\mathrm{i}\,A^{m+\beta}}{\Gamma\left(m+\beta\right)}a_{m}^{(0)} & 
\simeq &
-S_{1}\sum_{h=0}^{+\infty}a_{h}^{(1)}\prod_{\ell=1}^{h}\frac{A}{(m+\beta-\ell)}+\mathcal{O}\left(2^{-m}\right) 
\\ 
 & \simeq &
-S_{1}\left(a_{0}^{(1)}+ \frac{A}{m}a_{1}^{(1)} + \frac{A^{2}a_{2}^{(1)} -
  A\left(\beta-1\right)a_{1}^{(1)}}{m^{2}} + \cdots\right) +
\mathcal{O}\left(2^{-m}\right) \, .   \nn
\eea
To obtain the second line we expanded the first line for large $m$.
Now it becomes clear how one can determine the Stokes constant. 
Having calculated the coefficients $a_{m}^{(0)}$ and $a_{n}^{(1)}$
iteratively from the differential equation, we can now analyse the 
convergence of the lhs to the Stokes constant (times the value
of $a_{0}^{(1)}$), and thus determine $S_{1}$. 

\begin{figure}[ht]
\center
\includegraphics[height=0.4\textheight]{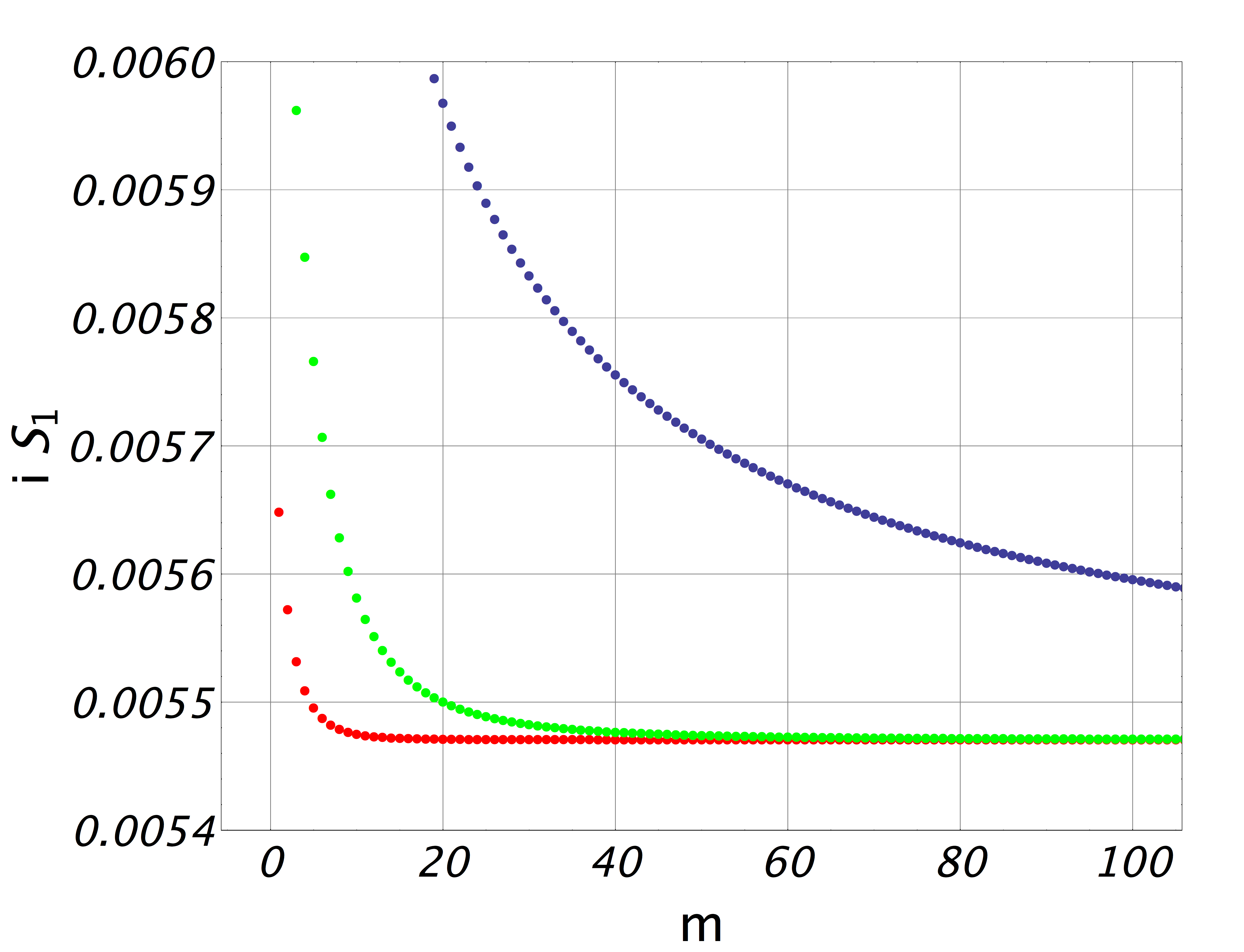}
\caption{Convergence of the large-order perturbative series (in blue) to the
  Stokes constant, using Richardson transforms of order 2 (green) and 5 (red). }
\label{fig:stokes-one-param}
\end{figure}

To carry out this calculation we make use of the accelerated convergence of  
the Richardson transforms \cite{Marino:2007te,Garoufalidis:2010ya,Schiappa:2013opa} (see 
Fig.~\ref{fig:stokes-one-param}). This leads to the determination of the
Stokes constant as $S_1=-0.00547029853 \,\mathrm{i}$, which matches the result
in Ref.~\cite{Heller:2015dha}.   

\begin{figure}[ht]
\center
\includegraphics[height=0.4\textheight]{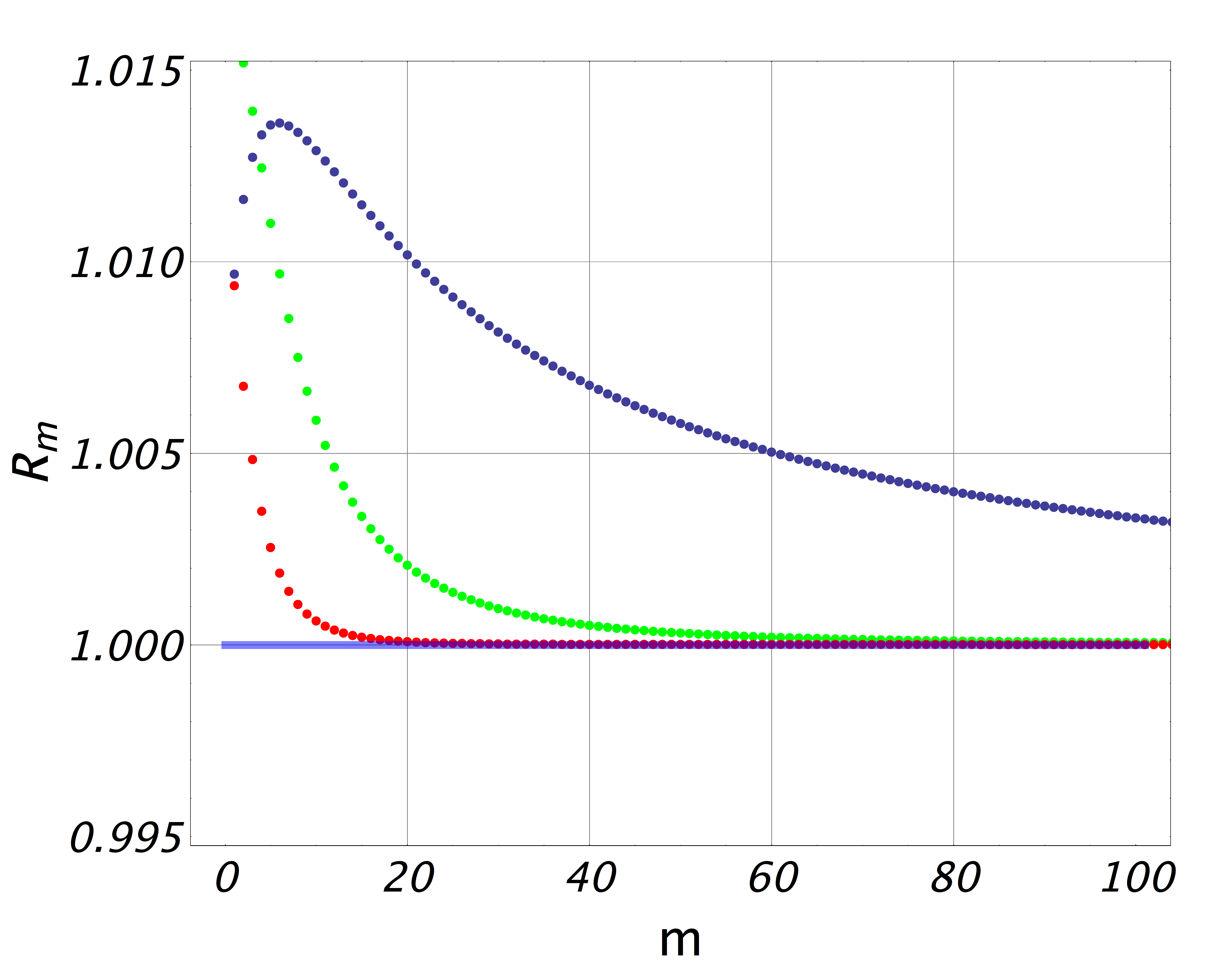}
\caption{Convergence of the ratio (\ref{eq:ratio-one-param-pert}), in blue, to
  the leading term in the large-order relation $c_0 = 1$ (light blue). The accelerated
  convergence is shown using Richardson transforms of 
  order 2 (green) and 5 (red).} 
\label{fig:ratio-one-param}
\end{figure}

It is important to note that we can check the
predictions obtained by resurgence techniques for the large-order
behaviour of the perturbative series even without the knowledge of
the Stokes constant. To do so, we analyse the convergence of the ratio 
of coefficients to the predicted values
\bel{rat1}
R_m  \equiv  \frac{a_{m+1}^{(0)}A}{a_{m}^{(0)}m} \, .
\ee
On the basis of the analysis we have presented above, we expect
\be
R_m \simeq\
 \left(1+\frac{\beta}{m}\right)\left(1-\frac{A}{m}\frac{a_{1}^{(1)}}{a_{0}^{(1)}}+\frac{A^{2}
   \left(a_{1}^{(1)}/a_{0}^{(1)}\right)^{2}+\left(2\beta-1\right)
   A\,a_{1}^{(1)}/a_{0}^{(1)}-2A^{2}a_{2}^{(1)}/a_{0}^{(1)}}{m^{3}} +
 \cdots\right) \, .
\ee
This quantity is clearly of the form
\be
R_m \simeq \sum_{k=0}^{+\infty} \frac{c_k}{m^k} \, ,
\ee 
where the coefficients $c_k$ are directly determined from the large $m$
expansion of the large-order relation given above. 
This makes it possible to use Richardson transforms to accelerate
  convergence. As seen in Fig.~\ref{fig:ratio-one-param}, the ratio
  \rf{rat1} converges to unity $c_{0}=1$ rather quickly. If the Richardson
  transform (of order 10) is used, already at $m=20$ this ratio  
differs from unity no more than one part in $10^8$ (and at $m=100$ the Richardson
transform of order 10 has an error of $10^{-16}$). 

\begin{figure}[ht]
\center
\includegraphics[height=0.4\textheight]{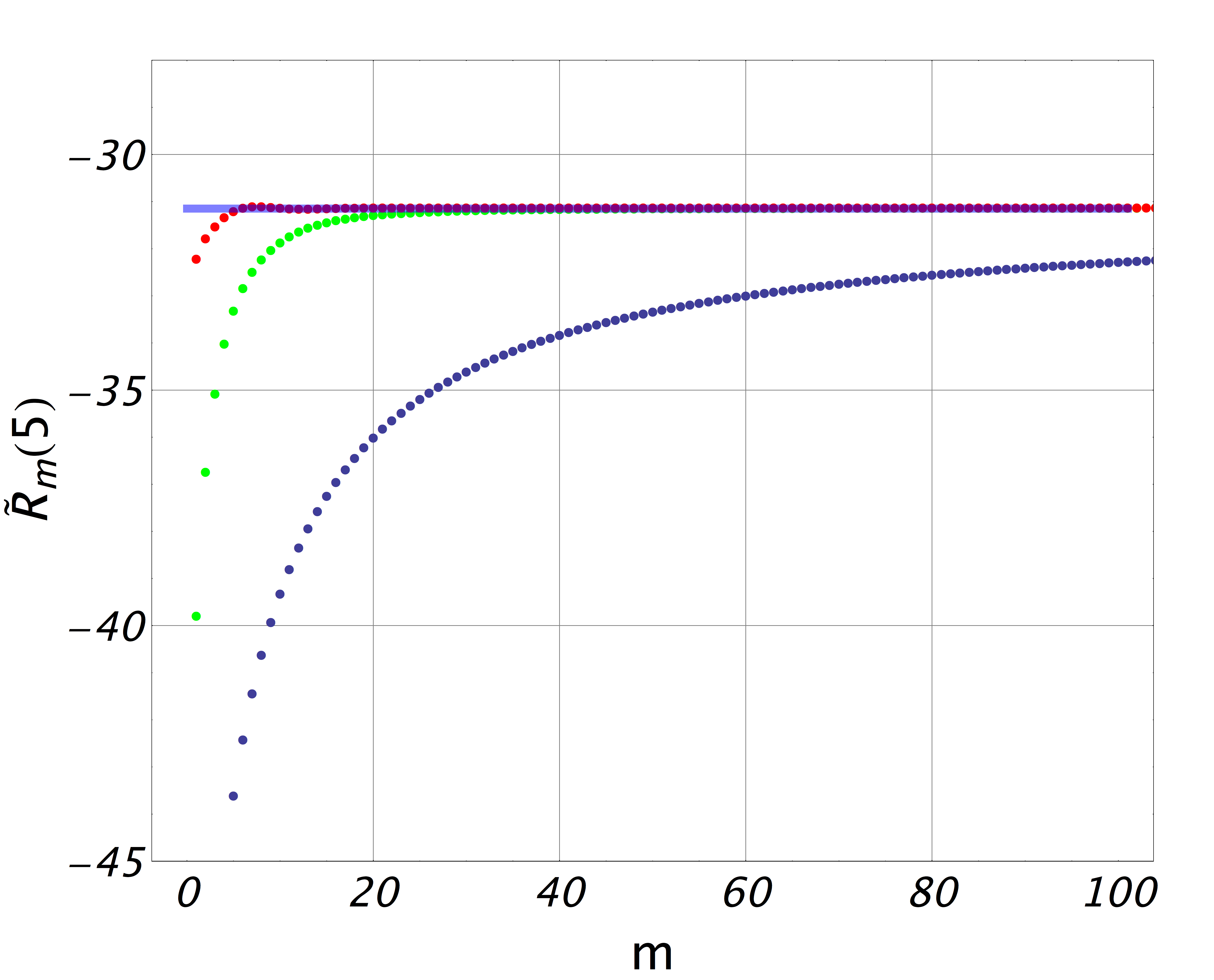}
\caption{Convergence of the ratio (\ref{eq:ratio-one-param-pert}), in blue, to
  the large-order relation coefficient $c_5$ (light blue). The accelerated
  convergence is shown using Richardson transforms of 
  order 2 (green) and 5 (red).} 
\label{fig:ratio-c5-one-param}
\end{figure}

Finally, note that one can
easily check for consistency the value of any coefficient $c_k$
predicted by resurgence by checking the convergence of  
\be
\widetilde{R}_m (k) \equiv \left(R_m-\sum_{r=0}^{k-1}\frac{c_r}{m^r}\right)\,
m^k \simeq c_k+\mathcal{O}(m^{-1}) \, .
\ee
In Fig. \ref{fig:ratio-c5-one-param} this convergence can be seen to the
predicted value of $c_5=-31.1456818997329$. 
For a Richardson transform of order 5, the error of the predicted value is
$10^{-8}$.  

This Section has served as a warmup for some of the resurgence
  techniques that we will be using in what follows. We needed to resolve a
  nonperturbative ambiguity due to a singularity on the positive real axis 
  and obtained results consistent with Ref.~\cite{Heller:2015dha}. The 
following Section applies resurgence techniques to an extended
  hydrodynamic theory which involves a second-order differential equation with two
Stokes lines in the complex plane, both away from the positive real axis. Even
though there will be no ambiguities in the resummation procedure
along the positive real axis, resurgence plays a determinant role in the
construction of the full transseries answer. The methods of resurgence that
will be used to 
study this problem are a generalisation of what was presented in this Section
and in Ref.~\cite{Aniceto:2011nu} (a review of nonlinear resurgent transseries
with two and more parameters can be found in Ref.~\cite{Upcoming:2015}).

Note finally that one can of course solve \rf{eq:first-order-MIS}
numerically. These exact solutions 
rapidly converge to an attractor which determines universal behaviour which
emerges after nonhydrodynamic degrees of freedom decay. The transseries
solution captures this behaviour: adding just the lowest order in the
transseries makes it possible to match the attractor solution by choosing the
real part of the transseries parameter $\sigma$ appropriately
\cite{Heller:2015dha}. 

While this work was in preparation, the authors became aware of
Ref.~\cite{Basar:2015ava} which has some overlap with this Section.

\section{Extended theories of hydrodynamics}
\label{sec:ext}

As reviewed in Section \ref{sec:mis}, MIS theory contains a single purely
damped nonhydrodynamic mode, the presence of which is reflected in the
divergence of the gradient expansion. The occurrence of this mode is enough to
furnish a causal hydrodynamic 
theory close to local equilibrium. This theory has been very successful in
describing the evolution of quark-gluon plasma produced in heavy ion
collisions, beginning with proper times of less than a fermi/c. It has,
however, been noted by many authors
\cite{Chesler:2009cy,Heller:2011ju,Jankowski:2014lna} that the pressure
anisotropy at these early times is still very large, and the system is not
close to equilibrium. It is natural to suspect that the nonhydrodynamic MIS
mode not only regulates the causality and stability issues of
Navier-Stokes hydrodynamics, but contributes in a very nontrivial way to the
physical implications of this model. This provides strong motivation to try to
understand better the role of nonhydrodynamic modes, and
how they can be matched with a microscopic description. For modeling early
nonequilibrium dynamics one would expect that incorporating further
nonhydrodynamic degrees of freedom should provide a better description.

A significant step leading in this direction was taken in
Ref.~\cite{Heller:2014wfa}, where extended hydrodynamic theories were
formulated in 
the context of \symm. These theories attempted to match the effective theory
to the pattern of the least damped black brane quasinormal modes which govern
the approach to hydrodynamics. 

In this paper we focus on the simplest model discussed in
Ref.~\cite{Heller:2014wfa} in which there  
is a pair nonhydrodynamic modes which are not purely decaying. 
The familiar relaxation equation MIS theory, \rf{mis2}, is 
replaced by
\bel{eqpi2s}
 \left((\f{1}{T} \D)^2\right. + \left. 2\Omega_I \f{1}{T} \D + |\Omega|^2\right)
\Pi^{\mu \nu} = - \eta  |\Omega|^2 \sigma^{\mu\nu}  - c_\sigma \f{1}{T} \D\left(\eta
\sigma^{\mu\nu}\right) + \ldots \, ,
\ee
where the ellipsis denotes contributions of second and higher order in
gradients. The parameter 
\bel{omega}
\Omega \equiv \Omega_R + i \Omega_I
\ee
is the complex ``quasinormal mode'' frequency. The coefficient $c_\sigma$ affects
the region of stability in parameter space 
\cite{Heller:2014wfa}. By solving \rf{eqpi2s} in the gradient
expansion one can also check that $c_\sigma$ contributes to second order
transport coefficients. 
However, in our work this coefficient does not play a qualitative role
and we will set it to zero. 

The appearance of the second derivative in \rf{eqpi2s} is what leads to 
nonhydrodynamic modes which are not purely decaying.
Indeed, the linearization of equations \rf{cons} and \rf{eqpi2s} around flat
space reveals a pair of 
nonhydrodynamic modes with complex frequencies $\Omega$ and $-\bar{\Omega}$. In
the case of \symm\ the leading quasinormal mode frequencies have the values
\cite{Nunez:2003eq} 
\bel{omegas}
\Omega_{R} \approx 9.800 , \qquad  \Omega_{I} \approx 8.629 \, .
\ee
and these are the values we assume in our calculations\footnote{The values in
  \rf{omegas} differ from those in 
  Table 1 of Ref.~\cite{Nunez:2003eq} (corresponding to an operator of conformal
  weight $\Delta=4$) by a factor of $2 \pi$.}.  

As in the case of MIS theory, upon imposing boost invariance the hydrodynamic
equations reduce to an ordinary differential equation for the
temperature. 
Of course, in the present case, the equation is of third order.
Introducing new variables as in \rf{wfdef} one can rewrite it as a
second-order differential equation for the function $f(w)$: 
\bel{vcp}
w f^2 f''+\alpha  f f'+12 f^2 f'+w f f'^2+\frac{\beta +
\gamma  f + \delta f^2 + 12 f^3}{w} = 0 \, ,
\label{eq:bifex}
\ee
where $f'$and $f''$ are the first and second derivatives of $f(w)$
with respect to $w$, and
\bea
\alpha &\equiv& -8 + 2 w \Omega_I , \nn\\
\beta  &\equiv& -\f{128}{27} - \f{32}{27} C\eta C_{\tau\Pi} - \f{4}{9} w (C_\eta |\Omega|^2 -
8 \Omega_I) - \f{2}{3} w^2 |\Omega|^2  , \nn\\
\gamma &\equiv&   \f{176}{9}  + \f{4}{3} C_\eta C_{\tau\Pi}  - \f{32}{3} w \Omega_I +  w^2 |\Omega|^2, \nn\\
\delta &\equiv&  - \f{80}{3}  + 8 w \Omega_I  \, .
\eea
This is the analog of \rf{eq:first-order-MIS} of MIS theory.  

For physical reasons it is clear that at late times (large $w$) the solution
must tend to $2/3$, 
which corresponds to ideal fluid behaviour. It is easy to see 
  analytically that this is indeed the asymptotic solution. One can easily
  determine the large-$w$ expansion of solutions:
\be
f(w) =  \f{2}{3} +  \f{4 C_\eta}{9} \f{1}{w} + \f{8 C_\eta (C_{\tau\Pi}  + 2
  \Omega_I)}{27 |\Omega|^2}  \f{1}{w^2} + \dots  \, .
\ee
As expected, the first two terms coincide with what one obtains in
MIS theory (see \rf{eq:mishydro}), whereas the third term is
different. This series can be calculated 
up to essentially any order and can be shown to be divergent, as discussed in
much detail in the following section.  

At early times, which correspond to small values of $w$, one finds a 
unique real
power series solution of the form  
\be
f(w) =  \f{8}{9} + \f{9 C_\eta |\Omega|^2- 8 \Omega_I}{3(20 + 9  C_\eta
  C_{\tau\Pi})} w + \dots \, .
\ee
By examining numerical solutions of \rf{eq:bifex} it is
clear that (similarly to the case of MIS theory) this is the small $w$
behaviour of an attractor solution 
valid in the entire range of $w$. 

\begin{figure}[ht]
\center
\includegraphics[height=0.4\textheight]{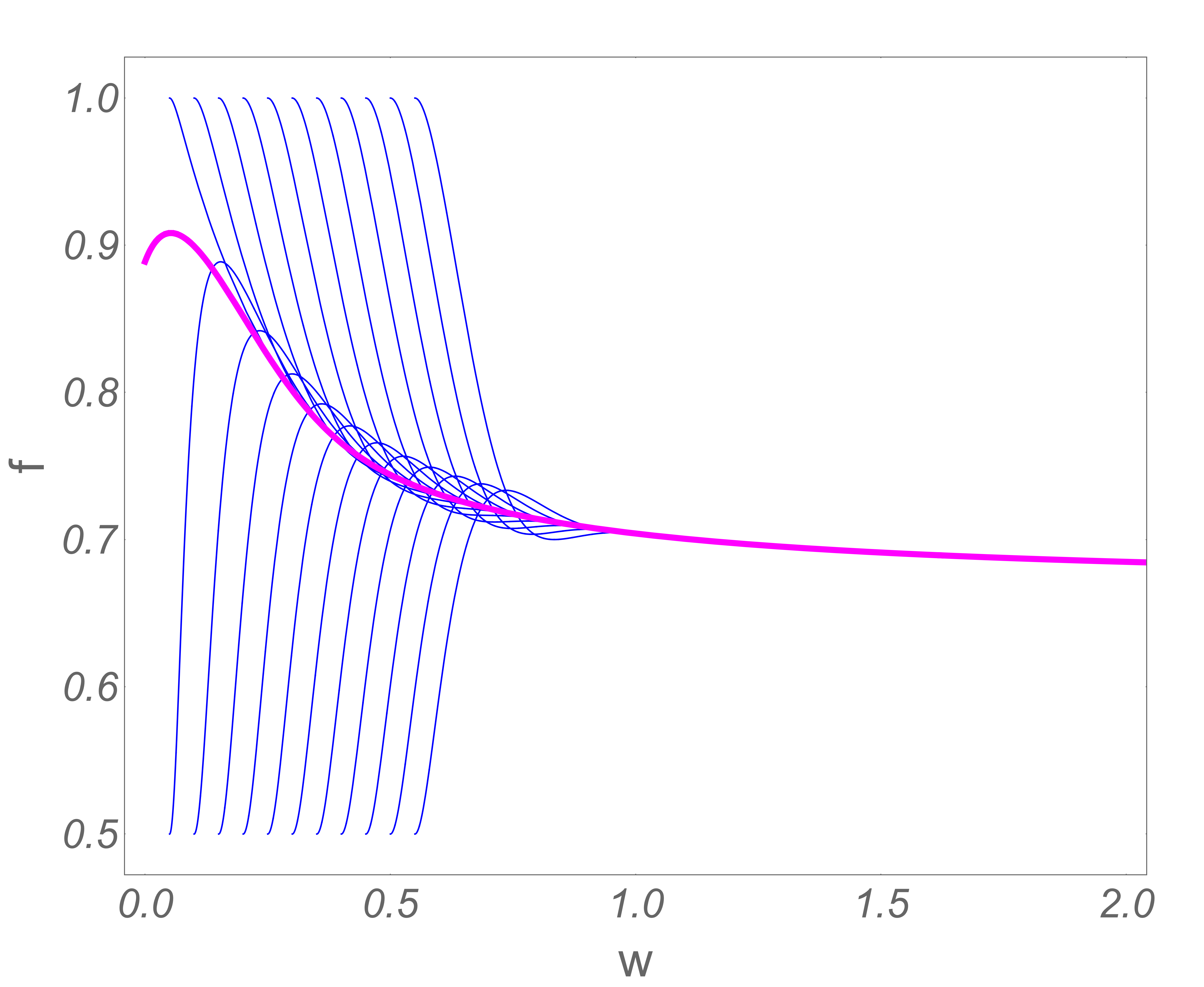}
\caption{Numerical solutions converge (nonmonotonically) to the numerical
  attractor (magenta). 
} 
\label{fig:attract}
\end{figure}

Since \rf{vcp} is of second order, one must
specify both $f$ and $f'$ at the initial value of $w$. 
As seen in Fig.~\ref{fig:attract}, setting
initial conditions at various values of $w$ shows that the numerical solutions
converge to the attractor.  
However, unlike in the MIS case, the numerical solutions do not decay monotonically but
oscillate around the attractor. 

\section{Resurgence in extended hydrodynamics}
\label{sec:extres}

We are interested
in solving \rf{eq:bifex} as an expansion in large values of $w\gg1$.
If we use a transseries Ansatz of the type (\ref{eq:trans-series-one-param})
and substitute it into this equation, we easily find two complex conjugate
values for the instanton action 
\be
A_{\pm}=\frac{3}{2}\left(\Omega_{I}\pm\mathrm{i}\Omega_{R}\right) \, ,
\ee
Equivalently, we can write $A_{+}=\frac{3}{2}\mathrm{i}\bar{\Omega}$,
while $A_{-}=-\frac{3}{2}\mathrm{i}\Omega$. We then have two types
of nonperturbative contributions and thus, following
Refs.~\cite{Aniceto:2011nu,Upcoming:2015},  
we find that we need a two-parameter transseries to fully describe
the solutions to this equation, 
\begin{equation}
f\left(w,\sigma_{\pm}\right)=\sum_{n,m=0}^{+\infty}\sigma_{+}^{n}\sigma_{-}^{m}
\mathrm{e}^{-(nA_{+}+mA_{-})w}\Phi_{(n|m)}\left(w\right) \, ,
\label{eq:trans-series-two-param} 
\end{equation}
where $\Phi_{(n|m)}(w)$ are the perturbative expansions in $w^{-1}$
around each sector. The perturbative sector is given by taking $n=m=0$.
These expansions are of the form 
\begin{equation}
\Phi_{(n|m)}\left(w\right)=w^{\beta_{n,m}}\sum_{k=0}^{+\infty}a_{k}^{(n|m)}w^{-k}
\, ,
\label{eq:exp-sectors-two-param}
\end{equation}
where the coefficients $\beta_{n,m}$ 
reflect the type of branch cut
singularities in the Borel plane, 
and the $a_{k}^{(n|m)}$ are the expansion
coefficients which 
can be determined iteratively by substituting Eq.~(\ref{eq:trans-series-two-param})
into Eq.~(\ref{eq:bifex}). Assuming furthermore
that $\beta_{n,m}=n\beta_{+}+m\beta_{-}$, we find 
\be
\beta_{\pm}=C_{\eta}\left(\Omega_{I}\pm\mathrm{i}\Omega_{R}\right) \, ,
\ee
together with recursion equations for the coefficients $a_{k}^{(n|m)}$.
Because $A_{\pm}$ are complex conjugate, as well as $\beta_{\pm}$,
and given that the coefficients of Eq.~(\ref{eq:bifex})
are all real, we 
see that that the coefficients $a_{k}^{(n|m)}$ will be complex conjugates 
of $a_{k}^{(m|n)}$ (and consequently all $a_{k}^{(n|n)}$ will be real).

\begin{figure}[ht]
\center
\includegraphics[height=0.4\textheight]{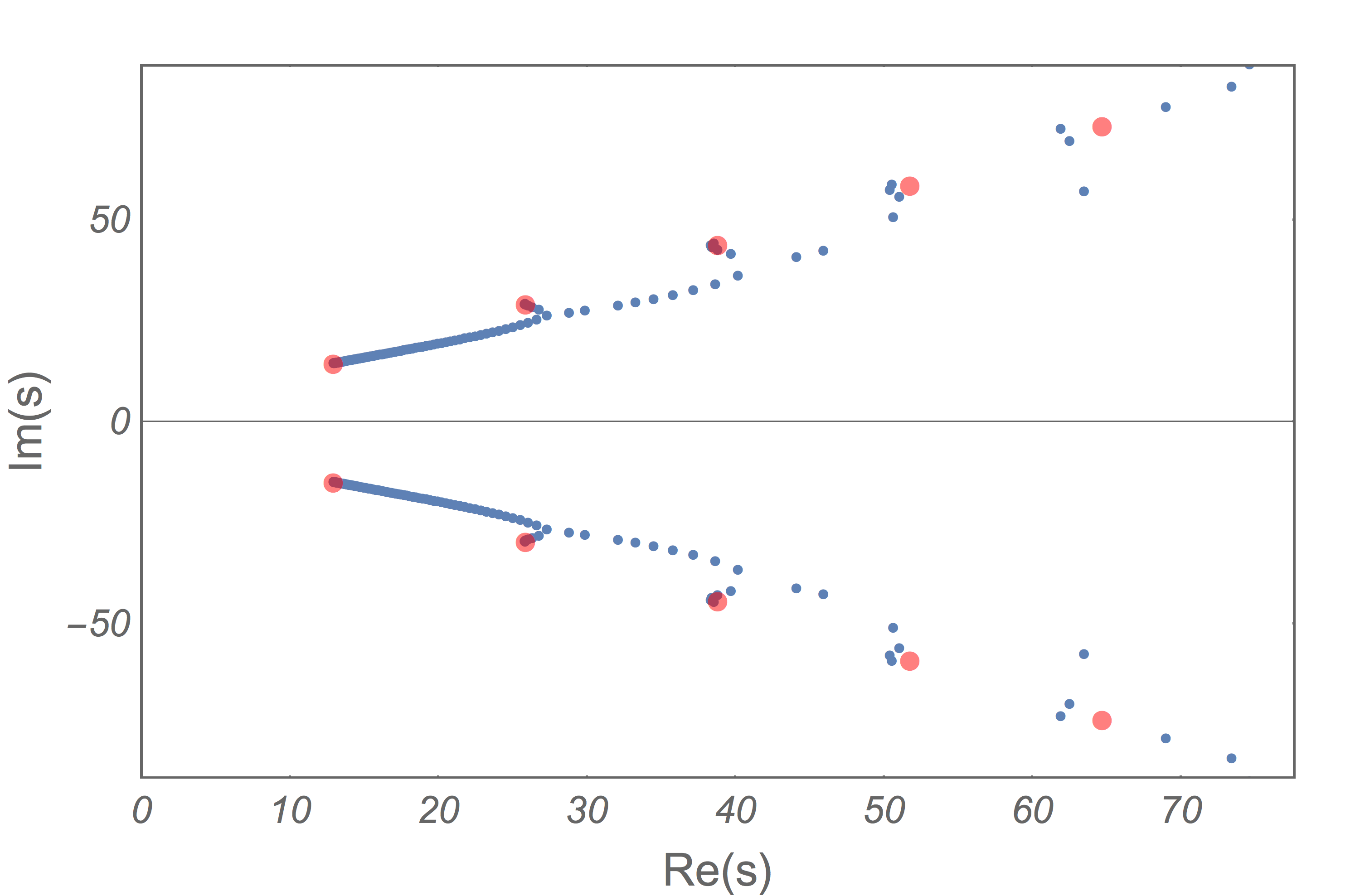}
\caption{Poles of the diagonal Borel-Pad\'{e}
approximant
of order 300 associated with $\Phi_{(0|0)}$ in the Borel $s-$plane.  
The red dots indicate values of the multiple instanton actions $\ell \,
A_{\pm},\,\ell\ge1$. 
  }  
\label{fig:Borelpoles}
\end{figure}

By studying numerically the behaviour of the coefficients of the perturbative series
we see that these grow factorially for large enough order $k$. This is
directly related to the behaviour of the Borel transforms. 
If we define the Borel transform for each sector 
\be
\mathcal{B}\left[\Phi_{(n|m)}\right]\left(s\right)=\sum_{k=k_{\mathrm{min}}}a_{k}^{(n|m)}\frac{s^{k-\beta_{n,m}-1}}{\Gamma\left(k-\beta_{n,m}\right)}
\, , 
\ee
we find a nonzero radius of convergence and branch cuts starting 
at positions $s_{\pm,\ell}=\ell \, A_{\pm}$. Note that $k_{\mathrm{min}}$ is
the minimum value of $k$ such that every power of $s$ appearing in the Borel
transform is non-negative (this does not change the asymptotic properties of
the series). In Fig.~\ref{fig:Borelpoles}, we see this behaviour 
for the Borel transform of the perturbative series
\be
\mathcal{B}\left[\Phi_{(0|0)}\right]\left(s\right) = \sum_{k=0}a_{k+1}^{(0|0)}
\frac{s^{k}}{\Gamma\left(k+1\right)} \, .
\ee

In order to analyse the singularities of the Borel transform, we use the
method of Pad\'{e} approximants, where the series above is approximated by a
ratio of polynomials.\footnote{In the diagonal case used here, this ratio is
  of the same order, half of the number of coefficients determined for the
  original series (see for example Ref.~\cite{Aniceto:2011nu} for more details).} 
Positions of the zeros of the polynomial in the denominator 
reflect the singular behaviour of the Borel transform: these poles condense in
certain directions, and indicate cuts in the Borel plane.

We check the resurgent properties of the transseries
\rf{eq:trans-series-two-param} by determining the large-order behaviour
predicted by resurgence for the perturbative series $\Phi_{(0|0)}$ (the
procedure can then 
be generalised for higher sectors). We first need to determine the associated
discontinuities, and then 
make use of Cauchy's theorem (\ref{eq:Cauchy-thm}), in the same manner
as for the one-parameter example previously studied.
\footnote{Note that for higher sectors there will naturally be additional
  singular directions in the Borel plane, associated with different
  combinations of the two instanton actions $A_{\pm}$, much in the same way as
  for the one-parameter transseries the sectors $\Phi_n,\,n\ge2$ has
  discontinuities in both the positive and negative real axis.} 

\begin{figure}[ht]
\center
\includegraphics[height=0.4\textheight]{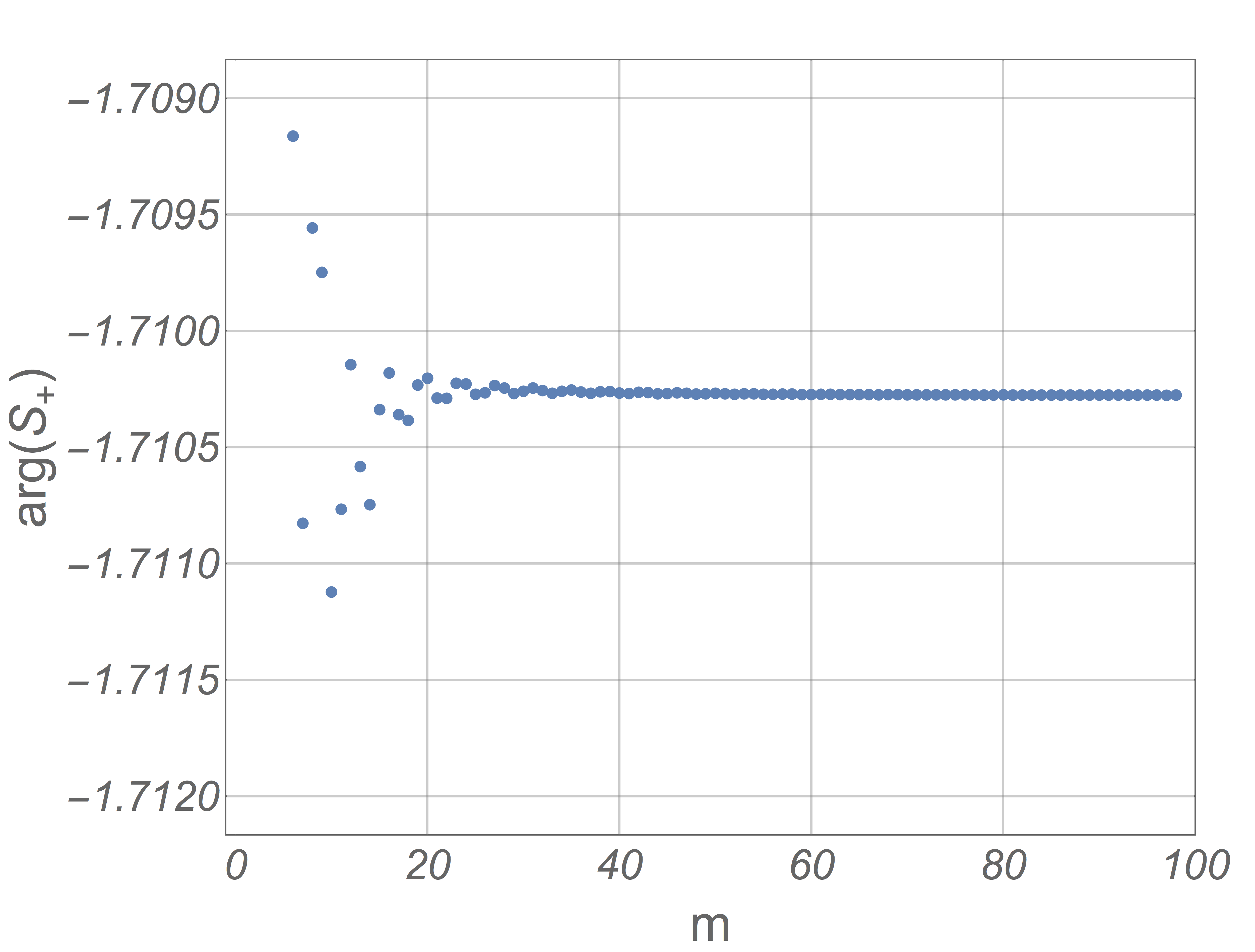}
\caption{Convergence of the phase of the Stokes constant $\theta_S \equiv
  \arg\left(S_{+}\right)$. We plot  (\ref{eq:large-ord-argS-v2}) for each $m$
  using the resummed $Q_m$. 
} 
\label{fig:stokesarg}
\end{figure}

In the present case we have two singular directions defined by the
two actions $A_{\pm}$:
\be
\theta_{\pm}=\pm\arctan\left(\frac{\Omega_{R}}{\Omega_{I}}\right)\equiv\pm\theta_{A}.
\ee
 Each of these directions will have a different Stokes constant associated
with it, which we will call $S_{\pm}$. 
Following the ideas of \cite{Aniceto:2011nu,Upcoming:2015}
we can write down the discontinuities associated with the singular directions as 
\begin{eqnarray*}
\mathrm{Disc}_{\theta_{+}}\Phi_{(0|0)}\left(w\right) & = &
-\sum_{\ell=1}^{+\infty}\left(S_{+}\right)^{\ell}\mathrm{e}^{-\ell
  A_{+}w}\Phi_{(\ell|0)}\left(w\right),\\ 
\mathrm{Disc}_{\theta_{-}}\Phi_{(0|0)}\left(w\right) & = &
-\sum_{\ell=1}^{+\infty}\left(S_{-}\right)^{\ell}\mathrm{e}^{-\ell
  A_{-}w}\Phi_{(0|\ell)}\left(w\right) \, .
\end{eqnarray*}
Rewriting these results for the variable $x=w^{-1}$,
making use of Cauchy's theorem (\ref{eq:Cauchy-thm}) for the function
$x\Phi_{(0|0)}\left(x\right)$, and expanding for small
$x$, we arrive at the large-order predictions ($m\gg1$)
\bea
a_{m}^{(0|0)} & \simeq &
-\sum_{k\ge1}\frac{\left(S_{+}\right)^{k}}{2\pi\mathrm{i}}\frac{\Gamma\left(m+k\beta_{+}\right)}{\left(kA_{+}\right)^{m+k\beta_{+}}}\sum_{h\ge0}a_{h}^{(k|0)}\frac{\Gamma\left(m+k\beta_{+}-h\right)}{\Gamma\left(m-k\beta_{+}\right)}\left(kA_{+}\right)^{h}- 
\nn\\  
 &  &
-\sum_{k\ge1}\frac{\left(S_{-}\right)^{k}}{2\pi\mathrm{i}}\frac{\Gamma\left(m+k\beta_{-}\right)}{\left(kA_{-}\right)^{m+k\beta_{-}}}\sum_{h\ge0}a_{h}^{(0|k)}\frac{\Gamma\left(m+k\beta_{-}-h\right)}{\Gamma\left(m-k\beta_{-}\right)}\left(kA_{-}\right)^{h} \, .
\eea
Given that all the coefficients $a_{m}^{(0|0)}$ are real, and that
the pairs $\beta_{\pm},\,A_{\pm}$ and $a_{h}^{(k|0)},a_{h}^{(0|k)}$
are complex conjugate, we can easily see that $\frac{S_{+}}{2\pi\mathrm{i}}$
has to be complex conjugate of $\frac{S_{-}}{2\pi\mathrm{i}}$, and
so the Stokes constants are related by 
\be
S_{-}=-\overline{S}_{+} \, .
\ee
It will be convenient to define 
$A_{\pm}=\left|A\right|\mathrm{e}^{\pm\mathrm{i}\theta_{A}}$, 
$\beta_{\pm}=\left|\beta\right|\mathrm{e}^{\pm\mathrm{i}\theta_{\beta}}=\beta_{R}\pm\mathrm{i}\beta_{I}$,
$S_{\pm}=\pm\left|S\right|\mathrm{e}^{\pm\mathrm{i}\theta_{S}}$. 
The leading 
behaviour of the large-order relations written above is 
dictated by the sectors $a_{h}^{(1|0)}$ and $a_{h}^{(0|1)}$, 
\be
a_{m}^{(0|0)} \simeq -\frac{S_{+}}{2\pi\mathrm{i}}
\frac{\Gamma\left(m+\beta_{+}\right)}{A_{+}^{m+\beta_{+}}}
\sum_{h\ge0}a_{h}^{(1|0)} \prod_{\ell=1}^{h}
\frac{A_{+}}{(m+\beta_{+}-\ell)}+h.c.+\mathcal{O} \left(2^{-m}\right)  \, , 
\label{eq:large-order-pert}
\ee
where $h.c.$ stands for the Hermitian conjugate. Unlike the one-parameter case
previously studied, in these large-order relations there will always be a
dependence on the Stokes constant $S_{+}$. Thus, before proceeding with deeper
tests of these relations, we need to numerically determine the Stokes
constants. This can be done as follows. Defining
\be
Q_m = - \frac{1}{2\pi\mathrm{i}} \frac{\Gamma\left(m+\beta_{+}\right)}{A_{+}^{m+\beta_{+}}}
\sum_{h\ge0}a_{h}^{(1|0)} \prod_{\ell=1}^{h} \frac{A_{+}}{(m+\beta_{+}-\ell)}
\label{eq:def-Qm}
\ee
one has 
\be
a_{m}^{(0|0)} \simeq S_{+} Q_m + \mathrm{h.c.}+\mathcal{O}\left(2^{-m}\right)
\, .
\label{eq:conv-pert-series-Qm}
\ee
If we can determine a resummed value for the $Q_m\equiv
\left|Q_m\right|\mathrm{e}^{\mathrm{i}\theta_{Q}(m)}$, for each $m$, then it
will easily follow that 
\be
\f{a_{m+1}^{(0|0)}}{a_{m}^{(0|0)}} \simeq \f{|Q_{m+1}|}{|Q_{m}|}
\f{\cos\left(\theta_Q (m+1) +\theta_S\right)}{\cos\left(\theta_Q (m) +
  \theta_S\right)} \, . 
\label{eq:large-ord-argS-v1}
\ee
Note that this relation is still a large-order relation, i.e., it is
valid for large values of $m$. The argument of the Stokes constant can then be
found by rewriting this large-order relation \rf{eq:large-ord-argS-v1}, 
\be
\tan\theta_S = \frac{g(m) \cos\theta_Q(m) -  \cos\theta_Q(m+1)}{g(m)
  \sin\theta_Q(m) -  \sin\theta_Q(m+1)} \, ,  
  \label{eq:large-ord-argS-v2}
\ee
where
\be
g(m)\equiv \f{a_{m+1}^{(0|0)}}{a_{m}^{(0|0)}} \f{|Q_m|}{|Q_{m+1}|} \, .
\ee
To determine the resummed values of $Q_m$, first notice that the sum present
in (\ref{eq:def-Qm}) is asymptotic for large $m$: 
\be
\eta(m)\equiv\sum_{h\ge0}a_{h}^{(1|0)} \prod_{\ell=1}^{h}
\frac{A_{+}}{(m+\beta_{+}-\ell)} \, \simeq  \,
\sum_{k=0}^{+\infty}\,\frac{\eta_k}{m^k} \, .
\label{eq:asympt-large-m}
\ee
The coefficients $\eta_k$ are fully determined by the value of
$A_{+},\,\beta_{+}$ and the coefficients $a_h^{(1|0)}$. The latter were 
determined from the recurrence relations coming from the original differential
equation, up to $h=100$. The above sum can be computed via the
Borel-Pad\'{e} resummation method (see Ref.~\cite{Aniceto:2011nu} for more
details):\footnote{This sum could also be approximated by performing an
  optimal truncation for each value of $m$.}  
\begin{itemize}
\item We first determine the Borel transform corresponding to the asymptotic
  sum $\eta(m)$, \rf{eq:asympt-large-m}.
\item We approximate this Borel transform by a diagonal Pad\'{e}
  approximant of order $N=50$, denoted by $\mathrm{BP}_{50}\left[\eta\right]$.  
\item The resummed series $\mathcal{S}\eta (m)$ is then determined via the
  usual Laplace transform.
along the positive real axis as we want $m\in\mathbb{N}$ 
\be
\mathcal{S}\eta(m) \,= \,\int_{0}^{+\infty}
ds\,\mathrm{e}^{-s\, m}\,\mathrm{BP}_{50}\left[\eta\right]\left(s\right) \, .
\ee
This was performed for
$m=1,\cdots,100$.\footnote{For 
    this sum, the positive real axis is 
    not a Stokes line, and there is no ambiguity associated with the
    resummation.}  
\end{itemize}
We can finally rewrite the resummed $Q_m$ as
\be
\mathcal{S}Q_m = - \frac{1}{2\pi\mathrm{i}} \frac{\Gamma\left(m+\beta_{+}\right)}{A_{+}^{m+\beta_{+}}}
\mathcal{S}\eta(m) \, .
\label{eq:resummed-Qm}
\ee
With this result we can determine the argument of the Stokes constant
$\theta_S\equiv\arg\left(S_{+}\right)$ for each value of $m$ via the relation
(\ref{eq:large-ord-argS-v2}), by substituting 
the resummed value $\mathcal{S}Q_m$ for the $Q_m$. 
The result is illustrated in
Fig. \ref{fig:stokesarg}: the phase becomes essentially independent of $m$ and
is given by $\theta_S = -1.710276$ (the estimated error is of order
$10^{-6}$).  

\begin{figure}[ht]
\center
\includegraphics[height=0.4\textheight]{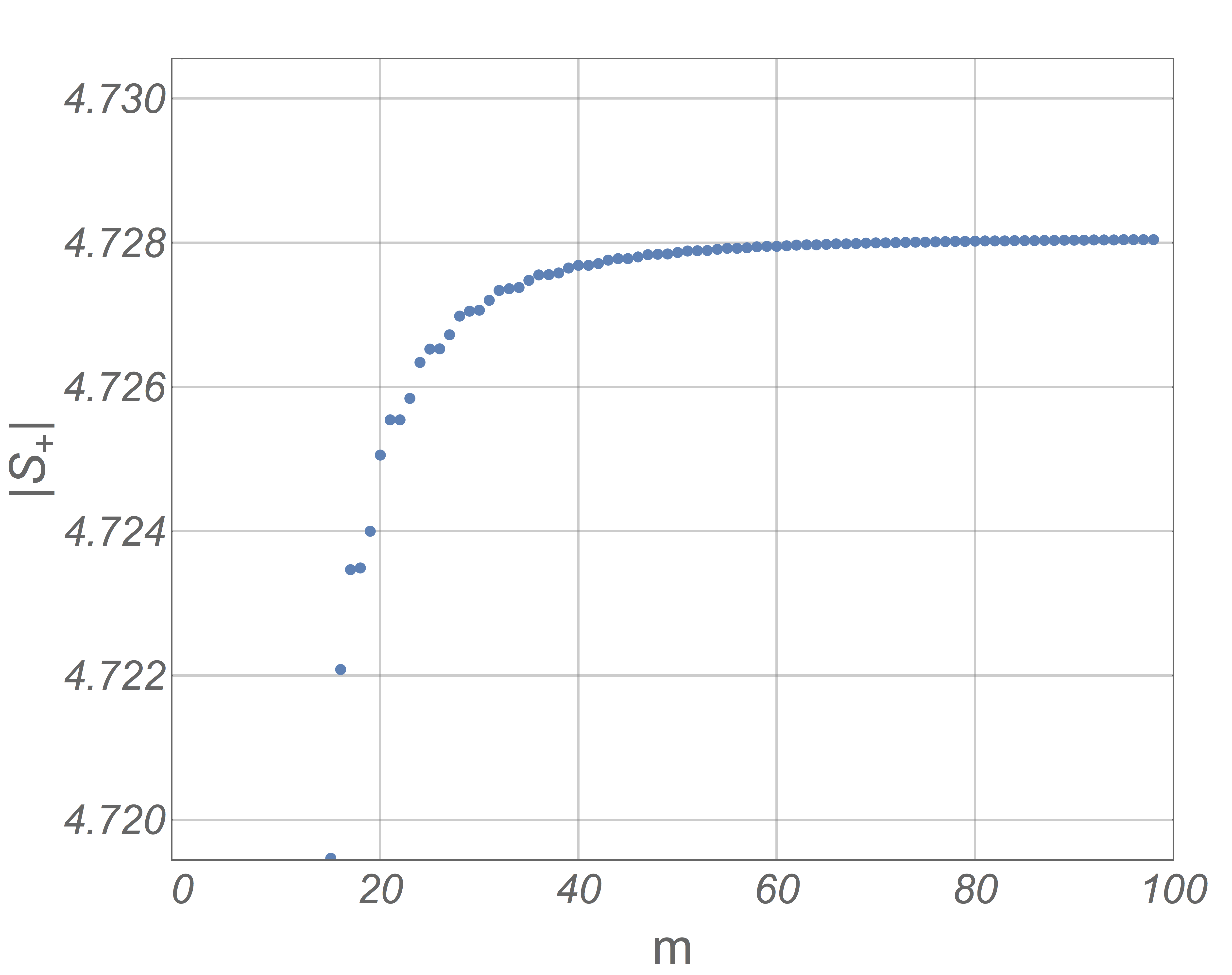}
\caption{Convergence of the large-order relation Eq.~(\ref{eq:large-ord-modS})
  to the modulus of the Stokes constant $\left|S\right| \equiv
  \left|S_{+}\right|$.}   
\label{fig:stokesmod}
\end{figure}

One can similarly calculate the modulus of the Stokes constant. 
From \rf{eq:conv-pert-series-Qm} it follows that for large $m$ 
the modulus of the Stokes constant $\left|S\right|\equiv\left|S_{+}\right|$  
should converge to 
\be
\left|S\right|\simeq\frac{a_m^{(0|0)}}{2\left|Q_m\right|\cos\left(\theta_S +
  \theta_Q (m)\right)} \, ,
\label{eq:large-ord-modS}
\ee
with the $Q_m$ 
replaced by the resummed ones given in \rf{eq:resummed-Qm}. 
This convergence can be seen in
Fig. \ref{fig:stokesmod}, 
implying the value $\left|S\right|=4.728045$ (with error of order $10^{-6}$). 

This concludes the numerical calculation of the Stokes constant. Using its
value we can now check the resurgence large-order relations
(\ref{eq:large-order-pert}). 

Let us define
a new quantity by 
\be
\Omega\left(m\right) \equiv \frac{2 \pi}{\left|S\right|}
\frac{\left|A\right|^{m+\beta_R} \mathrm{e}^{-\beta_{I}
    \theta_A}}{m^{\beta_R}\,\Gamma(m)} \, a_m^{(0|0)} \, .
\ee
Making use of the asymptotic expansion (\ref{eq:asympt-large-m}), as well as
the 
following large $m$
expansion\footnote{As an example we present the first
  three terms in this expansion: $\gamma_0=1$, $\gamma_1=\frac{1}{2}\,\beta_{+}\left(\beta_{+}-1\right)$, and
  $\gamma_3=\frac{1}{24}\,\beta_{+}\left(\beta_{+}-1\right)\left(\beta_{+}-2\right)\left(3\beta_{+}-1\right)$.}
\be
\gamma(m) \equiv
\frac{\Gamma(m+\beta_{+})}{m^{\beta_R}\,\Gamma(m)}=\sum_{k=0}^{+\infty}\frac{\gamma_k}{m^k}
\, ,
\ee
we can easily find the large order behaviour for $\Omega(m), m\gg1$:
\be
\Omega (m) \, \simeq \,\mathrm{e}^{\mathrm{i}\,\Theta (m)} \sum_{k=0}^{+\infty} \frac{c_k}{m^k}+h.c.
 \, = \, 2 \sum_{k=0}^{+\infty} \frac{\left|c_k\right|}{m^k} \,
 \cos\left(\Theta(m) + \theta_c(k)\right) \, .
 \label{eq:large-order-check}
\ee
The coefficients $c_k \equiv
\left|c_k\right|\mathrm{e}^{\mathrm{i}\theta_c (k)}$ appearing in this
expression are defined by  
\be
\sum_{k \ge 0} c_k\,m^{-k} \equiv \gamma(m) \, \eta(m)
\ee
and the angle $\Theta(m)$ is 
\be
\Theta (m) \equiv \frac{\pi}{2} + \theta_S - \theta_A
\left(m+\beta_R\right)-\beta_I \log\left|A\right| \, .
\ee
Note that the coefficients $c_k$ are known numerically, because both
expansions $\eta(m),\,\gamma(m)$ are known.  

\begin{figure}[ht]
\center
\includegraphics[height=0.4\textheight]{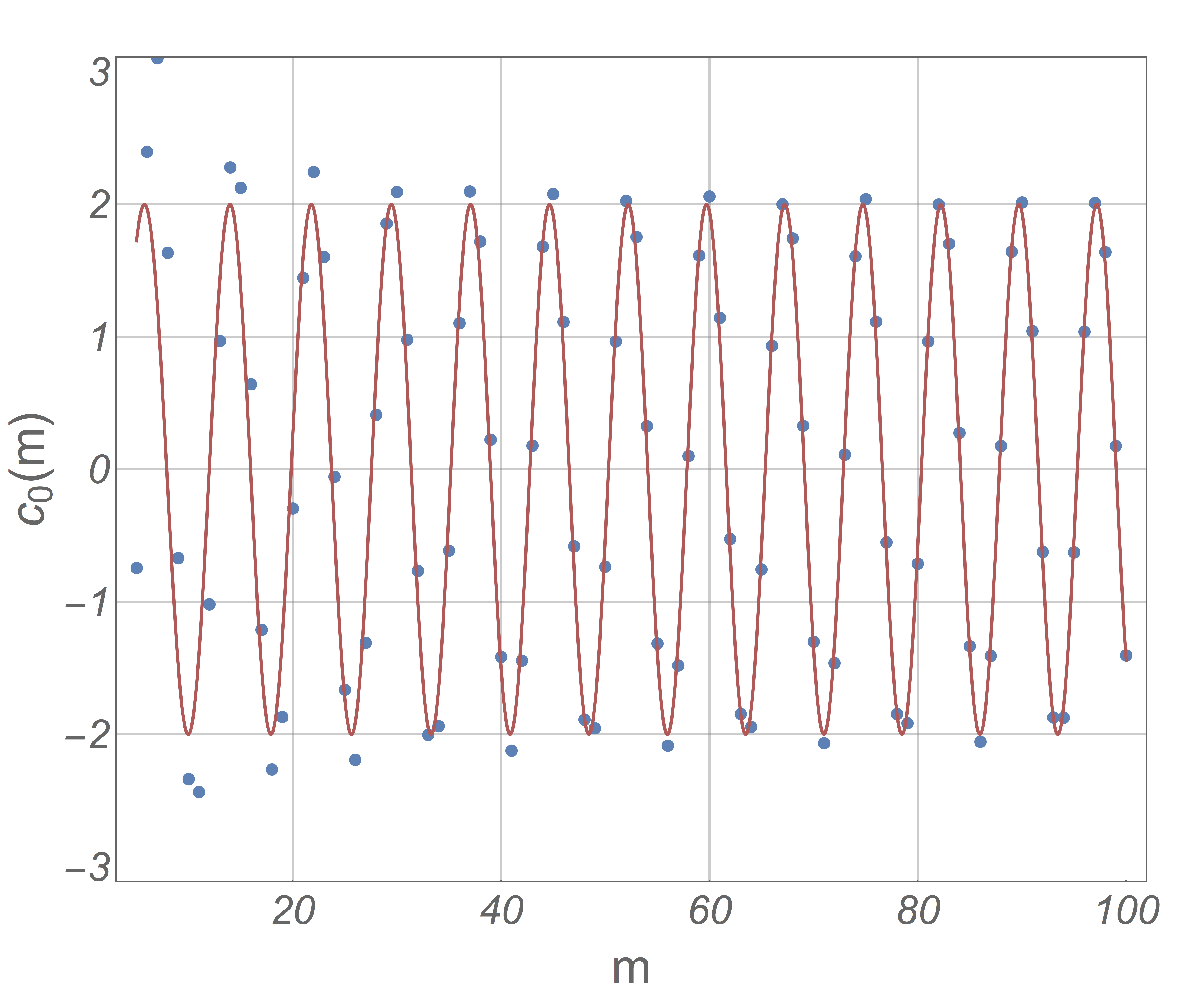}
\caption{Convergence of the large-order relation (\ref{eq:large-order-check})
  to the expected behaviour predicted by the coefficient $c_0(m)\equiv 2
  \cos\left(\Theta(m) \right) $.}  
\label{fig:large-ord-coeff0-5-100}
\end{figure}

\begin{figure}[ht]
\center
\includegraphics[height=0.4\textheight]{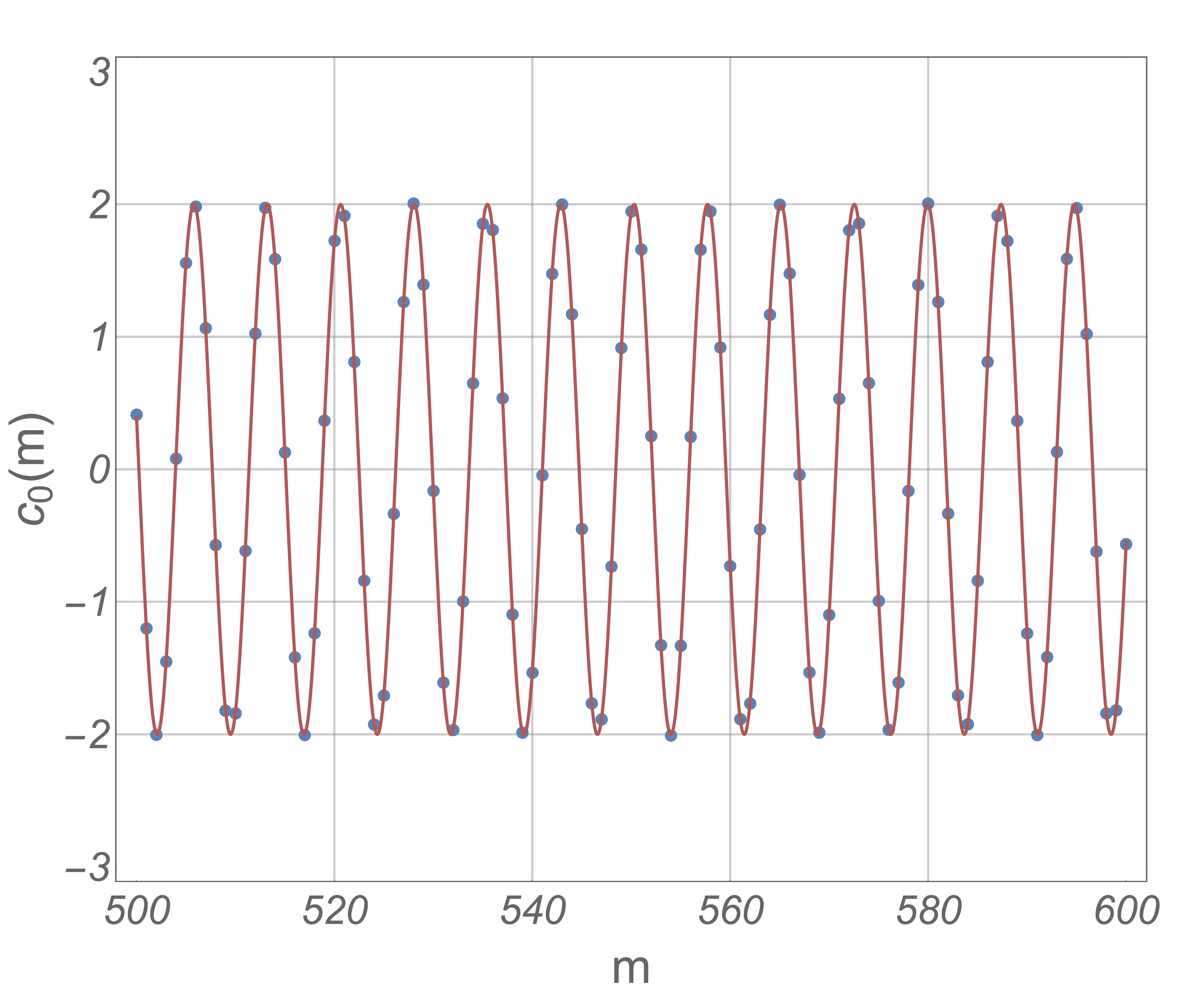}
\caption{
Consistency of the large-order relation (\ref{eq:large-order-check})
  with the expected behaviour predicted by the coefficient $c_0(m)\equiv 2
  \cos\left(\Theta(m) \right) $ at large $m$.
  }  
\label{fig:large-ord-coeff0}
\end{figure}

Analysing the relation (\ref{eq:large-order-check}) we find that to leading
order in $m$ (
since
$c_0=1$) 
\be
\Omega(m) \simeq 2 \cos\left(\Theta(m)\right) + \mathcal{O}\left(m^{-1}\right)
\, . 
\ee
In Figs. \ref{fig:large-ord-coeff0-5-100} and \ref{fig:large-ord-coeff0} we
can see the convergence of the numerical results to the predicted behaviour
for two different ranges of $m$:  for the range of $m<100$ we can see a slow
convergence, getting more accurate for higher values of $m$; in the range
$500<m<600$ there is already complete consistency between numerical results
and predicted behaviour. 

\begin{figure}[ht]
\center
\includegraphics[height=0.4\textheight]{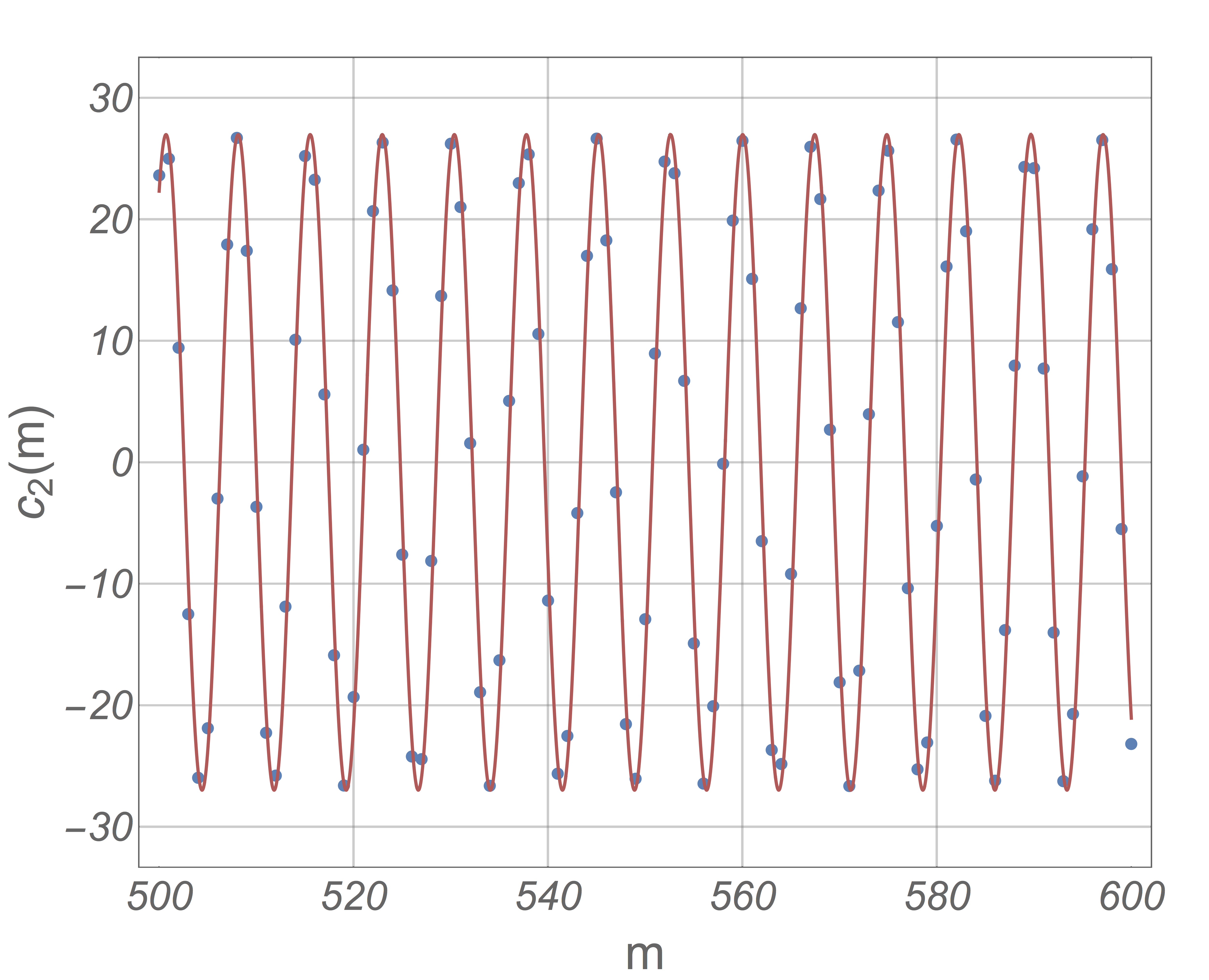}
\caption{
Consistency of the large-order relation
  \rf{eq:large-ord-coeff-c2} with the expected behaviour predicted by the
  coefficient $c_2(m)\equiv 2 
  \left|c_2\right| \, \cos\left(\Theta(m) + \theta_c(2)\right) $ at large $m$.
  }  
\label{fig:large-ord-coeff2}
\end{figure}

We can also study the convergence of the large-order relations to a general coefficient $c_k$ for some
specific $k$ by subtracting the first $k-1$ elements of the series and
multiplying the result by $m^k$.
 For example, we can check the convergence to
the term $k=2$ in the relation (\ref{eq:large-order-check}) by plotting 
\be
\left(\Omega(m) - 2 \sum_{k=0}^{1} \frac{\left|c_k\right|}{m^k} \,
\cos\left(\Theta(m) + \theta_c(k)\right)\right)m^2 \simeq  2 \left|c_2\right|
\, \cos\left(\Theta(m) + \theta_c(2)\right) +\mathcal{O}\left(m^{-1}\right). 
\label{eq:large-ord-coeff-c2}
\ee
This convergence can be seen in Fig. \ref{fig:large-ord-coeff2}
: for the range of large $m$ presented, we find consistency between numerical and predicted results.
It is important to note that the convergence to higher coefficients $c_k$ is
highly nontrivial, and is based on the assumption that the transseries is
resurgent 
and that the value of the Stokes constant 
has been correctly determined. If either of these assumptions had failed, 
we would not have found convergence of the numerics to higher 
orders predicted by resurgence. 

It is also of importance to point out the slight deviation of the numerical
data from the predicted values
in Fig. \ref{fig:large-ord-coeff2}. The main reason for this is that 
we have determined 
(based on the same numerical data) 
the values of the Stokes constants with an
  error of $10^{-6}$; this error will eventually cause such a
  deviation. In order to get more 
accurate results in the convergence to higher
  sectors, one would need to determine more coefficients of the sectors
  $\Phi_{(1|0)}$ and $\Phi_{(0|1)}$ and use them to lower the numerical error
  of the Stokes constant calculation. 

Now that we have confirmed the resurgent properties of the perturbative
series,\footnote{The resurgent properties of higher nonperturbative sectors
  can also be checked, once higher sectors are determined via recursion
  relations and resummations of the lower sectors are performed.} 
we turn to the central question: how to resum our two-parameter transseries 
(\ref{eq:trans-series-two-param}).
We want to resum our transseries for positive real coupling. Because 
the singularities in the Borel plane are away from this direction,
we can perform the integration of the Laplace transform (\ref{eq:resummation-one-param-sector}),
where now the sectors being resummed are the $\Phi_{(n|m)}$. There is no
ambiguity involved in this calculation, and the resummed transseries 
(for the positive real line $\theta=0$) is 
given by
\begin{equation}
\mathcal{S}_{0}f\left(w,\sigma_{\pm}\right) = \sum_{n,m=0}^{+\infty}
\sigma_{+}^{n} \sigma_{-}^{m}\mathrm{e}^{-(nA_{+}+mA_{-})w} \mathcal{S}_{0}
\Phi_{(n|m)} \left(w\right) \, .
\label{eq:resummed-trans-series-two-param}
\end{equation}
We could be tempted to set the transseries
parameters $\sigma_{\pm}=0$, which would leave us with only the perturbative
series. But the nonperturbative sectors will give us some real exponentially
suppressed contributions that we should not neglect as they will play
a role in the correct final answer.\footnote{If the actions had a negative real
  part, then one should in fact set 
the parameters to zero when considering the transseries in the positive
real axis, as not to have exponentially enhanced contributions.} In fact,
this was already seen in other problems of resummation
\cite{Grassi:2014cla,Couso-Santamaria:2015wga}. 
Consequently, we should allow for nonzero $\sigma_{\pm}$.

For real values of $w$ 
we expect 
a real solution, and we know that sectors $\Phi_{(n|m)}$ are complex
conjugate to $\Phi_{(m|m)}$, with the instanton actions also being
complex conjugates. Therefore 
in order to have a real solution we need to have 
$\sigma_{+}^{n}\sigma_{-}^{m}$ be complex conjugate
to $\sigma_{+}^{m}\sigma_{-}^{n}$ for any $m,n$. In particular putting
$m=0,n=1$ we find  
\be
\sigma_{+}=\overline{\sigma}_{-}\equiv\sigma \, .
\ee
Writing the first few terms of the resummed transseries
(\ref{eq:resummed-trans-series-two-param}), 
we have ($A_{\pm}=A_{R}\pm\mathrm{i}A_{I}$)
\begin{eqnarray*}
\mathcal{S}_{0}f(w,\sigma) & = &
\sum_{n=0}^{+\infty}\mathrm{e}^{-nA_{R}w} \sum_{m=0}^{n}\sigma^{n-m}
\bar{\sigma}^{m}\mathrm{e}^{-\mathrm{i}(n-2m)A_{I}w}\mathcal{S}_{0}\Phi_{(n|m)}\left(w\right)\\   
 & = &
\mathcal{S}_{0}\Phi_{(0|0)}\left(w\right)+\mathrm{e}^{-A_{R}w}2\,\mathrm{Re}\left(\sigma\mathrm{e}^{-\mathrm{i}A_{I}w}\mathcal{S}_{0}\Phi_{(1|0)}\left(w\right)\right)+\\ 
 &  &
+\mathrm{e}^{-2A_{R}w}\left[2\,\mathrm{Re}\left(\sigma^{2}\mathrm{e}^{-2\mathrm{i}A_{I}w}\mathcal{S}_{0}\Phi_{(2|0)}\left(w\right)\right)+\left|\sigma\right|^{2}\mathcal{S}_{0}\Phi_{(1|1)}\left(w\right)\right]+\mathcal{O}\left(\mathrm{e}^{-3A_{R}w}\right) \, .
\end{eqnarray*}

Note that the complex number $\sigma$ is not determined by the above
analysis. This freedom corresponds exactly to the two integration constants expected
for a solution of a second-order ordinary differential equation and can be
fixed by imposing suitable initial conditions.

\section{Summary and conclusions}
\label{sec:sum}

The equations of hydrodynamics constitute a physically well-motivated coarse
grained description of a wide range of phenomena. It has recently become clear
that they provide a new area of application for resurgence ideas. We have
tried to describe a mature version of these ideas in the context of MIS
theory, which provides the simplest example of an infinite hydrodynamic
series. This series is divergent in a way which encodes information about the
nonhydrodynamic mode present in MIS theory. 

The main point of this paper was to apply these methods to a hydrodynamic
model which aims to describe a richer spectrum of nonhydrodynamic modes,
inspired by what is seen in \symm. We have shown in some detail that also in
this theory the hydrodynamic solution is the leading term in a transseries
expansion. These results confirm general expectations concerning the nature of
gradient expansions \cite{Dunne}. They also provide an
interesting example of resurgent transseries, where the nonperturbative
sectors not only have the expected exponentially suppressed behaviour at late
times ($w\gg1$), but also an oscillatory one. This oscillatory behaviour will
become more pronounced in early times, when these sectors are no longer
suppressed -- even though there are no ambiguities in this problem, the full
transseries is still needed to account for this behaviour. From the point of
view of resurgence theory, this oscillatory behaviour also brought novel
features. Because the large-order relations cannot be disentangled from the
Stokes constants, and one cannot use normal convergence acceleration methods
due to the oscillations, we needed to introduce a Borel-Pad\'{e} resummation
of the first nonperturbative sectors to accurately determine both modulus and
argument of the Stokes constant. This then allowed us to check the large-order
relations with high accuracy. 

From a physical perspective, one would like to understand cases where the
series expansion is generated directly from some underlying microscopic
quantum theory, such as strongly coupled \symm. In this case there is an
infinite sequence of nonhydrodynamic modes corresponding to the black brane
quasinormal modes \cite{Kovtun:2005ev,Nunez:2003eq}. To include more than a
single pair of complex conjugate quasinormal modes would involve
multiparameter transseries, with each quasinormal mode defining (in principle)
a separate Stokes line.

In practice one would first aim at understanding the effects of the leading
modes -- those with the longest relaxation times. In the case of
boost-invariant flow in \symm, the hydrodynamic series has already been
computed to 
high order, and the calculation of at least a few terms of the 1-instanton
sector series is feasible. In conjunction with the methods developed in the
study presented here, this opens up the possibility of at least checking
consistency with resurgence in this very important case.

\vskip 2em

{\bf Acknowledgments:} we would like to thank Michał Heller 
and Ricardo Schiappa for useful comments
on the manuscript. 
I.A. was supported by the National Science Centre grant
2012/06/A/ST2/00396. 
M.S. was supported by the National Science Centre
grant 2012/07/B/ST2/03794.

\bibliographystyle{utphys}
\bibliography{ehydro}

\end{document}